\documentclass[fleqn,usenatbib]{mnras}

\usepackage{newtxtext,newtxmath}

\usepackage[T1]{fontenc}

\DeclareRobustCommand{\VAN}[3]{#2}
\let\VANthebibliography\thebibliography
\def\thebibliography{\DeclareRobustCommand{\VAN}[3]{##3}\VANthebibliography}

\usepackage{graphicx}	
\usepackage[subrefformat=parens]{subcaption} 
\usepackage{booktabs,multirow}
\usepackage{amsmath}	
\usepackage{orcidlink}

\newcommand{\sersic}{Sérsic}

\title[ReveaLLAGN 3: Revealing the IR SEDs of LLAGN]{ReveaLLAGN 3: Revealing the Infrared Spectral Energy Distributions of Low Luminosity AGN}

\author[D. Ohlson et al.]{David Ohlson$^{\orcidlink{0009-0004-9457-2495}}$,$^{1}$\thanks{E-mail: david.ohlson@utah.edu}
Anil C. Seth$^{\orcidlink{0000-0003-0248-5470}}$,$^{1}$
Kameron Goold$^{\orcidlink{0000-0002-7743-9906}}$,$^{1}$
Nischal Acharya$^{\orcidlink{0000-0001-9738-3594}}$,$^{2}$
Torsten B\"oker$^{\orcidlink{0000-0002-5666-7782}}$,$^{3}$
\newauthor
Antoine Dumont$^{\orcidlink{0000-0003-0234-3376}}$,$^{4}$
Ruby Duncan$^{\orcidlink{0009-0002-4406-7330}}$,$^{5,6}$
Michael Eracleous$^{\orcidlink{0000-0002-3719-940X}}$,$^{7}$
Anja Feldmeier-Krause$^{\orcidlink{0000-0002-0160-7221}}$,$^{8}$
\newauthor
Juan Antonio Fern\'{a}ndez-Ontiveros$^{\orcidlink{0000-0001-9490-899X}}$,$^{2}$
Elena Gallo$^{\orcidlink{0000-0001-5802-6041}}$,$^{9}$
Andy D. Goulding$^{\orcidlink{0000-0003-4700-663X}}$,$^{10}$
Jenny E. Greene$^{\orcidlink{0000-0002-5612-3427}}$,$^{10}$
\newauthor
Luis C. Ho$^{\orcidlink{0000-0001-6947-5846}}$,$^{11,12}$
Sera B. Markoff$^{\orcidlink{0000-0001-9564-0876}}$,$^{5,6,21}$
Mallory Molina$^{\orcidlink{0000-0001-8440-3613}}$,$^{13}$
Richard M. Plotkin$^{\orcidlink{0000-0002-7092-0326}}$,$^{14,15}$
\newauthor
Almudena Prieto$^{\orcidlink{0000-0002-3585-2639}}$,$^{16,17,18}$
Shobita Satyapal$^{\orcidlink{0000-0003-2277-2354}}$,$^{19}$
Glenn van de Ven$^{\orcidlink{0000-0003-4546-7731}}$,$^{8}$
and Jonelle L. Walsh$^{\orcidlink{0000-0002-1881-5908}}$$^{20}$
\\
Affiliations are listed at the end of the paper
}

\date{Accepted XXX. Received YYY; in original form ZZZ}

\pubyear{\the\year{}}

\begin{document}
\label{firstpage}
\pagerange{\pageref{firstpage}--\pageref{lastpage}}
\maketitle

\begin{abstract}
The spectral energy distributions of active galactic nuclei (AGN) provide valuable information on the emission mechanisms taking place in their inner regions. Here we use the James Webb Space Telescope to study the near and mid-infrared spectral energy distributions of seven nearby low-luminosity active galaxy nuclei from 1.7 to 24~$\mu$m. We cleanly separate the AGN spectrum by performing spatial fits of the two-dimensional images across all wavelengths. We then compare these clean unresolved spectra to aperture-based spectra and show that they effectively remove galaxy contamination. Comparing our infrared measurements to X-ray emission, we find that low-luminosity AGN have enhanced near-infrared emission relative to more luminous AGN, a trend that correlates with radio-loudness, while their mid-infrared-to-X-ray ratios are comparable. A majority of our targets show strong compact silicate emission suggesting the presence of warm dust on parsec scales in these systems.
We also quantify the near- and mid-infrared power-law slopes and colors of low-luminosity AGN. In the near-infrared we find AGN of all luminosities have similar spectral slopes and colors, however in the mid-infrared, we find low-luminosity AGN typically have steeper slopes and redder colors than more luminous AGN. Both the strong silicate emission and steep mid-infrared slopes disagree with existing models for low-luminosity AGN spectral energy distributions.  

\end{abstract}

\begin{keywords}
galaxies: active -- accretion -- infrared: galaxies -- techniques: imaging spectroscopy
\end{keywords}



\section{Introduction}
\label{sec:intro}

Rapidly accreting supermassive black holes create some of the brightest objects in the Universe, however, a majority of supermassive black holes lying at the centres of galaxies are much fainter due to much lower accretion rates \citep[e.g.][]{ho2009}. Relative to their rapidly accreting counterparts, these low-luminosity Active Galactic Nuclei (LLAGN) share a number of unique features. First, LLAGN show weaker ultraviolet (UV) emission and lack the "big blue bump" thermal continuum relative to bright AGN \citep{Ho1999,Maoz2007,Ho2008,Eracleous2010,FernandezOntiveros2023,Kang2025}. Second, they show very weak or frequently absent broad emission lines 
\citep[e.g.][]{Elitzur2014} and have both weak high-ionization lines and strong low-ionization lines \citep[e.g.][]{heckman1980,goold2026}. Third, LLAGN are frequently radio-loud, with strong compact radio emission \citep{nagar2000,Nagar2005,Baldi2023}.

These LLAGN phenomena can be explained by changes in the accretion structure that occur in systems accreting at rates somewhere below 1\% of the Eddington accretion limit\footnote{In this paper we define LLAGN as systems with Eddington ratios log($\rm L_{Bol}/L_{Edd}$)~$<-3$.}. At these lower accretion rates, the inner accretion disc becomes optically thin creating hot, radiatively inefficient accretion flows \citep[RIAFs; see reviews by][]{Abramowicz2013, yuan2014}. Several subclasses of RIAF models exist including advection dominated accretion flows (ADAFs), where material is advected directly on to the black hole \citep{Narayan1995,yuan2014}, and models where the inefficiency is directly related to outflows or convection of energy \citep[e.g.][]{Blandford1999, Quataert2000}. The weak UV emission and the absence of broad lines in LLAGN are naturally explained by RIAF models due to the truncation of the inner accretion disc \citep[e.g.][]{trump2011,Yu2011}. Modern general relativistic magneto-hydrodynamic simulations have enabled detailed numerical investigation of hot accretion flows and their application to data from the Event Horizon Telescope \citep[e.g.][]{Porth2019,EventHorizon2022} and to jet launching \citep[e.g.][]{Liska2018,Liska2020}. Synchrotron emission from jets provides a natural explanation for the power-law spectral energy distributions in LLAGN extending from the radio into the infrared \citep{markoff2008,FernandezOntiveros2019,FernandezOntiveros2023}.

A substantial amount of energy in all AGN is emitted in the infrared part of the spectrum. A tight relation is found between the X-ray emission and the emission at 12~$\mu$m in AGN at a wide range of luminosities \citep[e.g.][]{krabbe2001,gandhi2009, asmus2015} that likely reflects the balance between AGN heating and cooling \citep{spinoglio1989}. In the near-infrared at $\sim$2~$\mu$m, the correlation with X-ray is less tight, but still clearly found \citep{burtscher2015,muller-sanchez2018} with an apparent excess in LLAGN \citep{dumont2020}. The dominant emission source at infrared wavelengths for higher luminosity AGN is thought to be dust emission from the torus or a disc wind \citep{RamosAlmeida2017,Hoenig2019}, however, as noted above, jet synchrotron emission in LLAGN can also contribute significantly at these wavelengths \citep{markoff2008,Prieto2016,eht2021,FernandezOntiveros2023}. The relative contributions of these components remain unclear in LLAGN due to the challenge of separating the compact AGN emission from that of the underlying galaxy. In this context, the mid-infrared silicate features provide an important diagnostic of the dust surrounding the nucleus. While obscured Seyfert galaxies often show silicate absorption \citep{hatziminaoglou2015}, LLAGN display a similar behavior as type 1s, with silicate features frequently detected in emission \citep{Mason2012,mason2013}. However, previous studies based on \textit{Spitzer} spectroscopy lacked the angular resolution needed to confirm the nuclear origin of this emission and the sensitivity required to obtain high-S/N spectra for large LLAGN samples. Conversely, previous photometric measurements did not provide the spectral resolution needed to characterize the silicate features themselves in these nuclei.

To gain a better understanding of these relations, we need physical models to describe LLAGN SEDs. The SEDs of LLAGN have been modeled in the past, often under an assumption of their dominant emission mechanism.  A number of studies interpret the power-law spectra of LLAGN in the radio and infrared  in the context of jet models \citep{Yuan2002,markoff2008,lucchini2019,FernandezOntiveros2023}. ADAF-based models have also been used to interpret the spectra of LLAGN \citep{Yu2011,nemmen2014,lopez2024}, including in the context of the fundamental plane of black hole accretion \citep{merloni2003,falcke2004,Yuan2009}. \citet{nemmen2014} fit 12 LLAGN SEDs and found some are ADAF-dominated, while others are jet-dominated. \citet{lopez2024} created a \textsc{CIGALE} \citep{yang2020,yang2022} module for LLAGN that assumes an ADAF+Truncated Disc model using results of \citet{nemmen2014}. However, the SED fits and models mentioned lack clean or high-angular resolution spectra in the IR. Existing IR data (\textit{Spitzer}, \textit{Wide-field Infrared Survey Explorer} (\textit{WISE})) likely include flux contamination from surrounding stars and interstellar medium (ISM) of the host galaxy. With James Webb Space Telescope (\textit{JWST}), we can decompose the central engine from its surrounding host to get a clean IR continuum of LLAGN for the first time. Furthermore, by obtaining continuous spectra, we can directly test the model SEDs and examine the broad silicate emission features associated with the nuclear dust component.

Previous works that attempt to deconvolve and recover point sources from surrounding extended emission utilize the Python-based WebbPSF \citep{perrin2014} to simulate \textit{JWST}'s complex point spread function \citep[PSF,][]{gonzolez-martin2025, leist2024}. We note that WebbPSF simulations are known to deviate from observations. They do not account for detector error, do not account for wavefront error of the NIRSpec Integral Field Unit (IFU) path, and show a preferential axis along the slit axis at short wavelengths. Because we want to extract clean SEDs across both NIRSpec and MIRI IFU wavelengths, empirical point source observations are key for accurate PSF convolution, despite possible inconsistencies between the background noise of target observations and point sources.

The goal of the ReveaLLAGN project is to study LLAGN at high angular and spectral resolution in the infrared for the first time. The \textit{JWST} Cycle 1 project obtained broadband spectra for seven nearby LLAGN. Results on the emission line spectra have been presented in \citet{goold2024} and \citet{goold2026}, with the latter paper also providing an overview of the project. The nuclear spectra presented in those papers show evidence for galaxy contributions, with absorption lines visible in the near-infrared portions of most of the galaxies' nuclear aperture spectra.  Here, we focus on characterizing the SEDs through careful morphological separation of galaxy and AGN components. In \S\ref{sec:data} we describe our data and PSFs, in \S\ref{sec:fit} we outline our approach to fitting our 2D integral field data. We present our unresolved nuclear LLAGN spectra in \S\ref{sec:results}, and then compare their infrared fluxes to the X-ray emission in \S\ref{sec:irx}. We discuss silicate features, LLAGN SED slopes, and their location in infrared selection diagrams in \S\ref{sec:discussion}, and conclude in \S\ref{sec:conclusion}.

\section{Data} \label{sec:data}

The ReveaLLAGN survey collected \textit{JWST} NIRSpec and MIRI/MRS data for seven nearby LLAGN (Cycle 1 GO Proposal ID 2016, PI: A. Seth), providing IFU data covering a wavelength range from 1.66 to 27.9~$\mu$m. We use NIRSpec instrument configurations F170LP/G235H and F290LP/G395H. These have a 3\arcsec$\times$3\arcsec~FOV, 0.1$''$ spaxel size, and a gap in wavelength coverage due to a physical gap between detectors ($\sim2.4-2.5~\mu m$ for G235H and $\sim3.9 - 4.2~\mu m$ for G395H). MIRI/MRS data spans four separate IFUs, channels 1 through 4: Ch1 ($4.9-7.65~\mu m$), Ch2 ($7.51-11.7~\mu m$), Ch3 ($11.55-17.98~\mu m$), and Ch4 ($17.7-27.9~\mu m$). We refer throughout this paper to the two separate NIRSpec grating settings as "channels" as well, since their data cubes are treated in a similar way to the individual channels in MIRI.
The field of view and spaxel size both increase for each MIRI/MRS channel: Ch1 ($3.2''\times3.7''$, $0.13''$), Ch2 ($4.0'' \times 4.8''$, $0.17''$), Ch3 ($5.2'' \times 6.2''$, $0.20''$), Ch4 ($6.6'' \times 7.7''$, $0.35''$). Each MIRI/MRS channel has a slightly different angular rotation with respect to the sky coordinate system: Ch1 ($8.4^\circ$), Ch2 ($8.2^\circ$), Ch3 ($7.5^\circ$), Ch4 ($8.3^\circ$).

We process the uncalibrated MIRI/MRS data using the \textit{JWST} pipeline \citep{bushouse2022} v1.15.1 and calibration context 1293. For NIRSpec data, we use the \textit{JWST} pipeline v1.15.0, calibration context 1256 and apply \textsc{WICKED} \citep{dumont2025} to correct the wiggles found in \textit{JWST} IFU spectra \citep{law2023}, caused by an undersampled PSF in combination with curved spectral traces. The details of our observations and procedure are fully described in \cite{goold2026}, and we expand on relevant details in Appendix~\ref{appendix:pipeline}.

We apply the galaxy properties from \citet{goold2026} to all seven ReveaLLAGN targets, listed here in Table~\ref{tab:target_table}. We also follow the methods of \citet{goold2026} to calculate $\rm L_{Bol}/L_{Edd}$ for our targets and archival comparison samples used throughout the paper by calculating (1) Eddington luminosity using $\rm M_{BH}$ and (2) bolometric luminosity using the methods of \citet{duras2020}, detailed in Appendix~\ref{appendix:eddington}.

\begin{table*}
	\centering
	\caption{\centering ReveaLLAGN Target Properties}
	\label{tab:target_table}
	\begin{tabular}{l r@{\,±\,}l ccccc}
		\hline\hline
		\textbf{Galaxy} & 
        \multicolumn{2}{c}{\textbf{Distance}} & 
        \textbf{log(M$_{{\star}}$)} &
        \textbf{AGN Type} &
        \textbf{log(M$_\text{BH}$)} &
        \textbf{log(L$_\text{2-10keV}$)} &
        \textbf{log(L$_\text{Bol}$/L$_\text{Edd}$)}
        \\
        \\
         & \multicolumn{2}{c}{(Mpc)}& [M$_{\odot}$] & & [M$_{\odot}$] & [erg~s$^\text{-1}$] & 
         \\
		\hline
            NGC~1052  & 19.41 & 2.41 & 10.70 &  L1.9  & 8.82 & 41.52 & $-$4.21 \\
            NGC~4258  & 7.58 & 0.08 & 10.66 &  S1.9  & 7.60 & 40.99 & $-$3.53 \\
            NGC~4395  & 4.76 & 0.02 & 8.39 &  S1.8  & 5.60 & 39.93 & $-$2.58 \\
            M81    & 3.70 & 0.16 & 10.70 &  S1.5  & 7.81 & 40.25 & $-$4.47 \\
            M87    & 16.70 & 0.60 & 11.32 &  L2  & 9.81 & 40.66 & $-$6.06 \\
            M94    & 4.41 & 0.08 & 10.40 &  L2  & 6.83 & 38.46 & $-$5.28 \\
            Sombrero  & 9.55 & 0.31 & 11.21 &  L2  & 8.83 & 39.98 & $-$5.76 \\
		\hline
	\end{tabular}
    \\
    \vspace{1mm}
    \raggedright
    {\small \textbf{References} -- {\bf Distances}: NGC~1052 -- \citet{Tonry2001}; NGC~4258 -- \citet{Reid2019}; NGC~4395, M81, and M94 -- \citet{Karachentsev2004}{$^\dagger$} ; M87 -- \citet{Blakeslee2009}; Sombrero -- \citet{McQuinn2016}. {\bf Galaxy Mass}: From \citet{bi2020}. {\bf AGN Type}: From \citet{Ho1997}, {\bf BH Mass}: NGC~1052 is based on velocity dispersion \citet{Koss2022}, NGC~4258 -- \citet{Reid2019}; NGC~4395 -- \citet{denBrok2015}; M81 and M94 -- \citet{Kormendy2013}; M87 -- \citet{EHT2019}; Sombrero -- \citep{Jardel2011}. {\bf 2-10~keV X-ray Luminosity}: 2-10~keV luminosities for Sombrero, NGC~1052, NGC~4258, NGC~4395, M81, and M87 from \citet{asmus2015}, M94 from \citet{Gonzalez-martin2009}, using distances in this table in all cases. {\bf Eddington Ratio}: Calculated using $L_{\rm{bol}}$ and $L_{\rm{Edd}}$ from Equations~\ref{eq:ledd}, \ref{eq:Kx_Lx}, \ref{eq:lbol}.}
    \\
    $\dagger$ Updated values from \href{https://relay.sao.ru/lv/lvgdb/}{Catalog \& Atlas of the LV Galaxies}
\end{table*}

\subsection{Construction of PSFs}\label{subsec:psf}

We choose to utilize publicly available observations of unresolved point sources to achieve empirically-based PSF convolution and avoid known shortcomings of WebbPSF simulations. For this purpose, we utilize NIRSpec observation of 2MASS J17571324+6703409 (PID: 3399, PI: M. Perrin; 16 dithers) and MIRI/MRS observation of HD192163 (PID: 1031, PI: A. Labiano; 4 dithers). These are both \textit{JWST} IFU calibration targets used by \citet{goold2026} to calculate the 75\% encircled energy aperture radii of the \textit{JWST} PSF across NIRSpec and MIRI/MRS wavelengths. We use these wavelength-dependent apertures for nuclear extraction in \S\ref{subsec:specextract}. We process both point source observations following the methods described in \S\ref{sec:data} to maintain consistency between all IFU data. The use of IFU-aligned cubes ensures that the non-axisymmetric PSF structure is aligned between our PSF images and science images.

The HD192163 observation has widely separated dither positions that provide further radial coverage in one direction, thus allowing accurate flux measurements only in parts of the field. We exploit the point symmetry of the \textit{JWST} PSF \citep{rigby2023} to fill the corners by applying a meshgrid to rotate, mirror, and convolve the 2D image at each single-wavelength slice\footnote{For the remainder of the paper, we refer to an image at a single wavelength plane extracted from the reconstructed data cube as a "wavelength slice".}.

\begin{figure}

    \begin{subfigure}[b]{.49\linewidth}
        \centering
        \includegraphics[width=\linewidth]{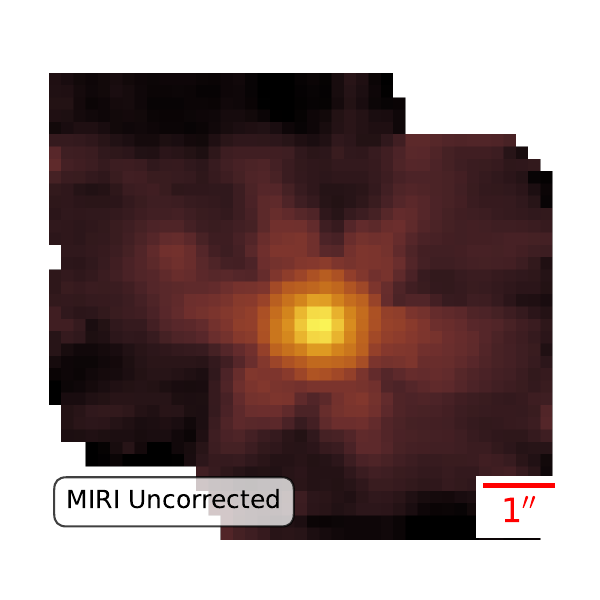}
        \caption{}
    \end{subfigure}
        \begin{subfigure}[b]{.49\linewidth}
        \centering
        \includegraphics[width=\linewidth]{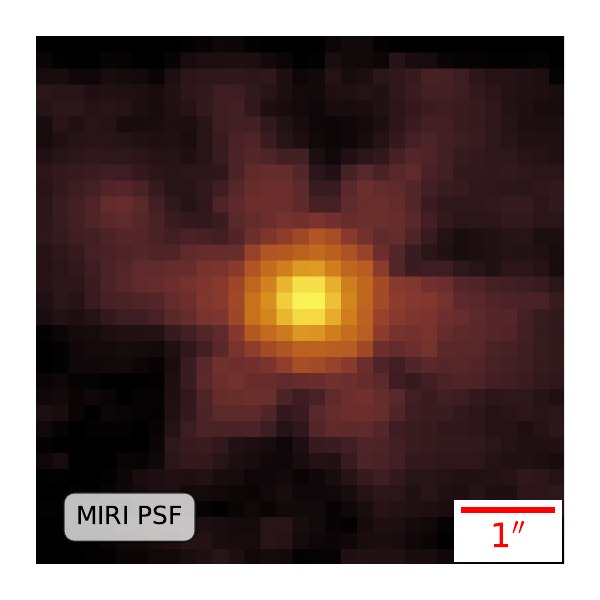}
        \caption{}
    \end{subfigure}%
    
    \centering
    \begin{subfigure}[b]{.49\linewidth}
        \centering
        \includegraphics[width=\linewidth]{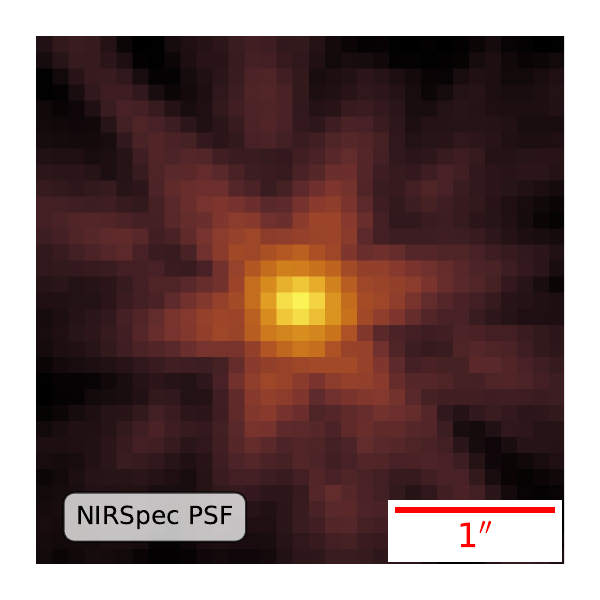}
        \caption{}\label{fig1a}
    \end{subfigure}\hfill
    
    \caption{Examples of the \textit{JWST} point source observations that we use for PSF convolution in our fitting process. All three figures depict a 2D image of the median flux across all wavelengths in a single channel. \emph{(a)} -- MIRI Channel 2 observations before corrections, which depict the corners of missing flux. \emph{(b)} -- MIRI Channel 2 used for fitting process. \emph{(c)} -- NIRSpec G395H.}
    \label{fig:psf}
\end{figure}

Our modeling method requires a centred PSF with a square field for convolution. We use the photometric centroid to select the central pixel of our PSF and negative pixels are replaced with a 2-pixel radial average at each wavelength slice. We then crop each channel data cube to an ZxNxN shape, where Z is the number of spectral pixels, and N is the largest pixel width relative to the PSF centre containing only valid flux values. Finally, we apply a moving median smoothing along the Z-axis (wavelength) to the resulting cube with a box width of $\pm$10 wavelength slices. This box width minimizes spectral noise without significantly altering the \textit{JWST} PSF's varying morphology. Because our chosen MIRI/MRS PSF is derived from observations of a Wolf-Rayet star, we also identified and applied wavelength median smoothing over significant emission features (which showed discontinuities in the PSF) to ensure a consistent empirical PSF across all MIRI/MRS channels. As an example, Fig.~\ref{fig:psf} shows (a) the Ch2 median image for HD192163 and (b) the corresponding channel median after the described process.

For the NIRSpec point source data, we spatially crop and spectrally smooth the processed cubes using the methods applied for our MIRI/MRS PSF cubes. The NIRSpec G395H median of our PSF model is shown in Fig.~\ref{fig:psf} (c).

\section{Methodology: Decontaminating AGN Spectra Through 2D Fitting} \label{sec:fit}

Extracting clean IR spectra of LLAGN requires accounting for contamination from the surrounding host galaxy. This galaxy contribution can include stellar light with absorption features and emission from gas surrounding the AGN that can skew the slope and shape of an observed SED. Accounting for the host contamination is particularly important for AGN with very low luminosities, a high inclination angle of the surrounding host, or a particularly dominant nearby stellar contribution.

Subtraction of a local background estimated from an annular aperture can potentially remove most of the host galaxy contamination in the central aperture. However, a host galaxy's brightness profile increases radially inward, potentially leading to undercorrection--especially in the NIR. In addition, the \textit{JWST} PSF is diffraction-limited longward of $\sim$1.1~$\mu m$, and thus has a width that increases linearly with wavelength. This needs to be taken into account in order to avoid affecting the SED shape.

To obtain the cleanest possible spectra for our sources, we use 2D surface brightness models fit with the package \textsc{Imfit} \citep{erwin2015}. \textsc{Imfit} is a 2D image modeler that fits an arbitrary number of user-specified functions convolved with a proved PSF using an iterative $\chi^{2}$ method. Chosen functions of the 2D fit can be combined as a single model and used separately as model components. Our method models the unresolved central engine as a point source and its surrounding host galaxy with a combination of flat-sky and elliptical \sersic{} profiles, while using the point-source observations described in \S\ref{subsec:psf} to model the \textit{JWST} PSF. We include additional components when residuals are high enough to warrant them.

In the following subsections, we describe the method used to separate the unresolved AGN component from the surrounding galaxy. We first fit initial models to the channel-wide median of an IFU data cube. We use the resulting fit to create a channel-wide template, with fixed morphology and variable intensity. Next, we use our template to fit 2D models at every wavelength slice in the data cube. Finally, we combine the resulting 2D models into model cubes and use these to extract spectra of both the central unresolved AGN and the surrounding galaxy light.

\subsection{Median Fitting}\label{subsec:medfit}

We start our modeling with templates fit to channel-wide images, calculated from a 2D image with the median flux values across all wavelengths within a single channel. Depending on the target, models contain a central point source, along with host galaxy components fit with a combination of flat sky, and one or two \sersic{} functions. We received clean MIRI/MRS data for our targets months before NIRSpec, so we started fitting with MIRI Channel 1 (4.9 - 7.65~$\mu m$). We continued building initial fit templates using Ch1 after incorporating NIRSpec data into the process, as its central wavelength range among our available spectra includes clear imaging of both AGN and galaxy components visible in our targets. Sombrero and M94 in particular had problems fitting to NIRSpec G235H data first; the dominant host galaxy components caused model degeneracy with the very dim point source, leading to noise and a generally underestimated AGN contribution.

To ensure these methods did not negatively affect fits of brighter point-like targets, we also attempted to base our model template morphology on an initial fit from a channel-wide median of NIRSpec G235H. This choice resulted in similar models, but a slightly lower average signal-to-noise ratio in 1D spectra and higher fractional residuals in MIRI imaging.

\begin{figure*}
     \centering 
     \includegraphics[width=0.475\linewidth]{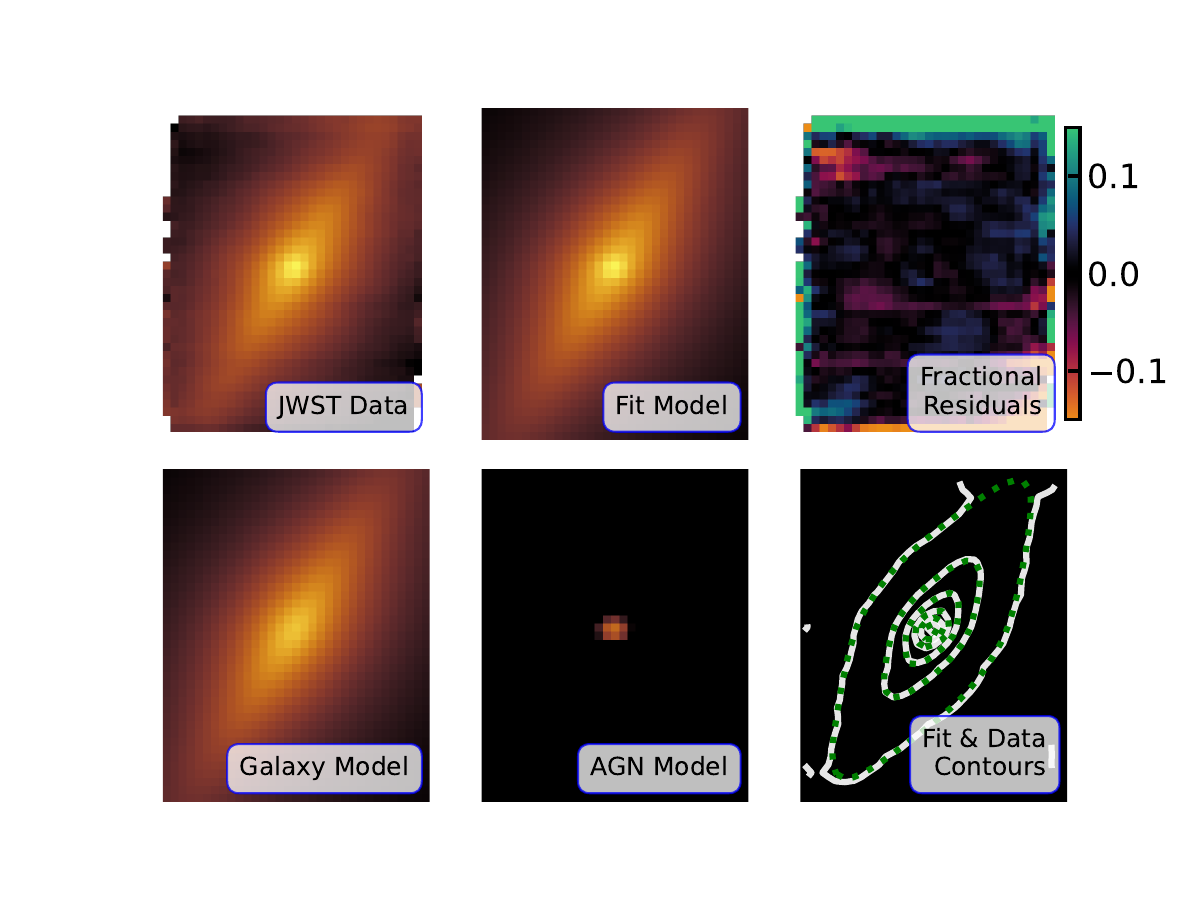}%
     \hfil 
     \includegraphics[width=0.475\linewidth]{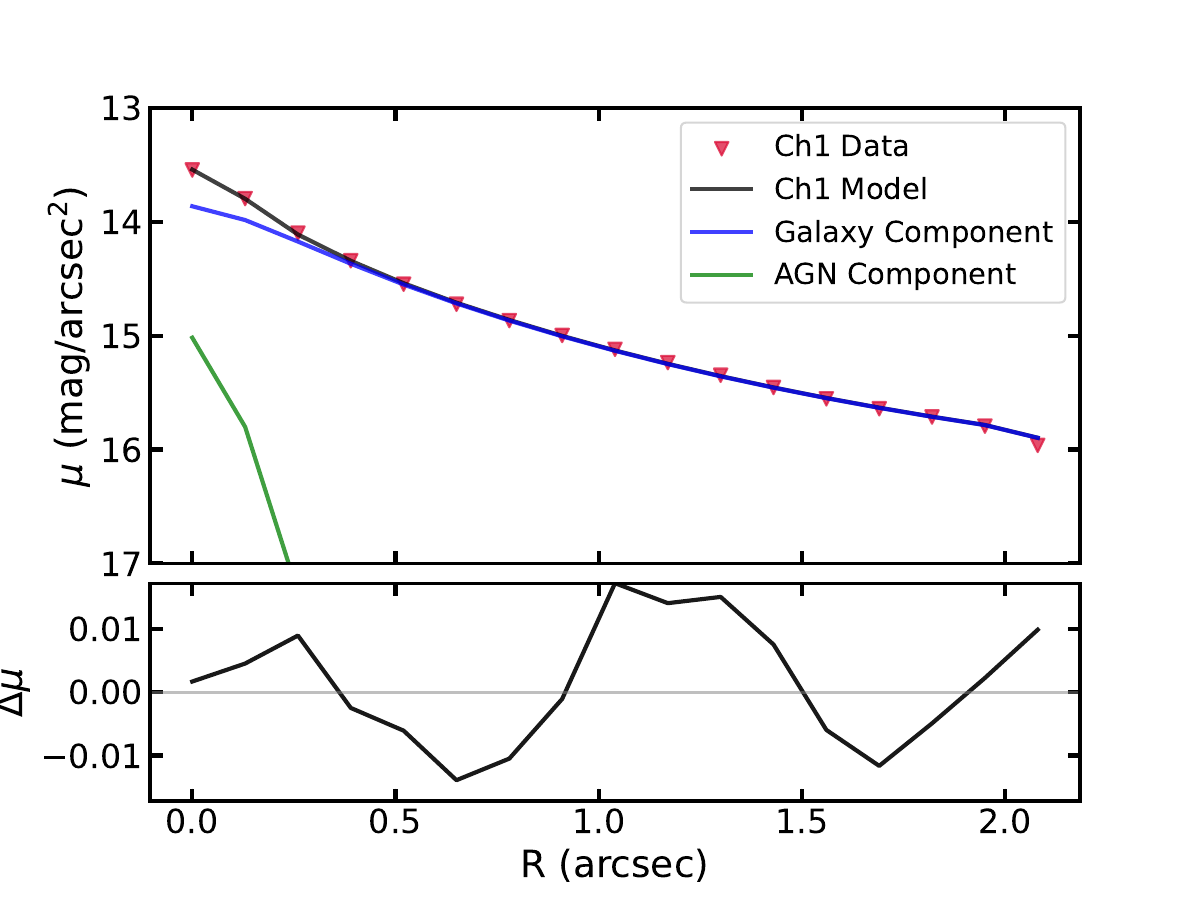}%
    \caption{Example of 2D fitting results using MIRI Ch1 channel-wide median of Sombrero. {\emph Left} -- 2D images depict the Ch1 data, our model, and the fractional difference between the two. Also shown are separated components of the surrounding host galaxy and unresolved AGN, and contours comparing the data (white) and our fit (green). {\emph Right} -- A comparison of radial brightness profiles of the \textit{JWST} Data, Fit Model, Galaxy Component, and AGN component, as well as the radial profile of fractional residuals. When creating the channel-wide templates, we individually inspected these figures to ensure they were satisfactory.}
    \label{fig:medfit}
\end{figure*}

We fit the channel-wide image for MIRI/MRS Ch1, carefully choosing initial values to minimize residuals between the model and the data and return an accurate radial profile, as shown in Fig.~\ref{fig:medfit}. The fit shown is typical, with individual pixels typically having $<$15\% residuals and radial profiles having $<$10\% residuals.
To maintain consistent modeling of stellar components, we use the MIRI Ch1 profiles as initial conditions for fitting other channels, allowing only small deviations as required to achieve good fits—transforming position angle and pixel scale when necessary. When applying \sersic{} functions from Ch1 fits to NIRSpec cubes, we allow $\pm0.1$ deviation to \sersic{} ellipticity for NGC~4258, NGC~1052, M81, M87, and M94. NGC~4395, M81, M94, and NGC~4258 required an additional \sersic{} profile for NIRSpec data. M87 is the only galaxy that required a changed \sersic{} index, $n$, when fitting to NIRSpec. We use the resulting channel-wide fitting results to create templates to model the components at each wavelength slice

\subsection{Iterative Fitting}\label{subsec:wavefit}

We use channel-wide templates described in \S\ref{subsec:medfit} to create a 2D model by fitting them to each wavelength slice of the channel data cube using fixed \sersic{} shape parameters and variable intensities. This method works well for obtaining clean continuum measurements of the unresolved central PSF, free from galaxy contamination; this measurement of the AGN flux is the central goal of this paper. However, the fits are less reliable when there are emission lines with extended flux, as the background morphology may differ significantly from our template model in these channels. This may affect the inferred fluxes from many emission lines; therefore, we recommend using aperture spectra for emission line measurements, as in \citet{goold2026}.

For each model at a single wavelength slice, we save separate 2D flux arrays of (i) the full model, (ii) model-data residuals, (iii) the Central Point Source/AGN, (iv) individual galaxy components (FlatSky, \sersic{}). We combine these together to generate spectral model and residual cubes, with one image per wavelength slice.

\subsection{Fitting for Special Cases} \label{subsec:specialgalaxies}

M94 and M87 require additional steps beyond the model variations described above to ensure accurate and realistic results.

\textbf{M94:} this nucleus shows the strongest flux contribution from its surrounding host galaxy. This bright galactic contribution exhibits a complex stellar light morphology that is poorly described by the Sersic profiles we use to model it. In addition, M94 also shows very strong polycyclic aromatic hydrocarbon (PAH) emission features across a significant range of wavelengths in the mid-infrared. This PAH emission has a separate, asymmetric morphology and is much brighter than the underlying stellar emission. To mitigate this second issue, we carefully investigated and selected ranges with strong PAH emission based on the central wavelengths reported by \citet{li2020}. We created two separate galaxy model templates: a stellar light template for wavelengths without strong PAH emission, which is also the template used to fit the NIRSpec data cubes, and a fit to the selected regions with strong PAH emission from the host galaxy.

Fitting the galaxy stellar template (without the influence of PAH emission) results in low residuals and good fits for near infrared and non-PAH impacted mid-infrared wavelengths. However, the model for PAH wavelengths was not fully successful. The two-model fitting method results in flux jumps at wavelengths where the fit transitions between models; this is discussed further in \S\ref{subsec:agnmodelspec}. We note that, because of the larger uncertainties in the PAH emission regions, we flag wavelengths where we applied the PAH template in the 1D spectra table of M94 and suggest caution when using the decomposed AGN spectrum in these ranges.

\textbf{M87:} the strong extended jet is visible throughout the near- and mid-infrared spectrum. The jet flux and morphology are prominent enough to skew residuals and influence accurate fitting, but \textsc{Imfit} does not include functions required to accurately model it. To remove jet emission and cleanly isolate the AGN point source spectrum, we masked spaxels with significant jet flux during the fitting process. Masked spaxels were chosen by eye, with consideration of whether their flux was significantly higher than that of others at the same radius. This adjustment allows our models to consider only the central AGN and host galaxy contributions. We also apply this mask to fractional residual maps at each wavelength to appropriately assess modeling accuracy.

\subsection{Spectral Extraction}\label{subsec:specextract}

As stated previously, we use our point-source model cubes to represent the unresolved AGN component in each target, and model the surrounding host galaxy emission by summing the remaining model component cubes. We extract our final 1D spectra from our 2D image models at each wavelength for both the point source and galaxy components using a circular aperture with fixed 1.7$''$ radius. This radius fits fully within the FOV of all channels. Using an oversized aperture at all wavelengths minimizes flux differences as a function of wavelength. This aperture size encompasses $\sim$90\% of point-source light at 24~$\mu$m, the longest wavelength we consider.

The formal uncertainties on our 1D spectra are very small due to the high S/N of the data, especially when considering the continuum across many pixels. This means our continuum SED fluxes are dominated by uncertainties due to modeling methods. We quantify those by using channel-median images and models to calculate a range of point-source intensity allowed for the given fit. We analyse fractional residuals of our model while varying the point source component flux by a factor between 0.1-5. We estimate the uncertainty on the AGN flux based on the level of model residuals in each channel. We do this by calculating the flux required to create a central profile residual equal to the typical fractional residual in the data cube, which we calculate using a robust estimator of the standard deviation $\sigma_{G}=0.7413(q_{75}-q_{25})$. This process allows us to carefully consider model-based flux errors of each IFU channel separately for all seven targets. Average flux errors from this method range from 13\% in NIRSpec G235H to 44\% in MIRI Channel 4; these errors are a good estimate of the possible systematic errors in our flux measurements. We assume flux errors calculated using this method for all pixels in the associated data cube and apply them to integrated flux and luminosity measurements in \S\ref{sec:irx} and \S\ref{sec:discussion}.

We also extract background-subtracted nuclear aperture spectra from our processed \textit{JWST} data cubes following the methods of \citet{goold2026} to test the quality of our AGN decontamination in \S\ref{subsec:comparespec} and quantify the strength and compactness of silicate emission in \S\ref{subsec:sil_measurement}.

Unresolved AGN component, galaxy component, and aperture-extracted spectra all show small flux discontinuities at borders between specific channels for all seven targets. Previous works have reported these flux discontinuities at the borders between subchannels, and used varying methods to stitch continuous 1D spectra \citep[e.g.][]{lai2025,ceci2025,canin2025}.
To correct these, we normalize the flux between neighboring channels' using their overlapping spectral pixels. In MIRI, channels 1-3 had minimal differences, and no corrections were applied. However, more significant discontinuities were seen between between Ch3 and Ch4. All galaxies except M94 had correction factors ranging from 0.93-1.18 ($\sim$2-18\% corrections), with M94 requiring a 35\% correction. We chose to normalize Ch4 to match Ch3 because of the noted noise and decreased fit quality in Ch4. We also see an offset between MIRI and NIRSpec; we choose to normalize the NIRSpec to MIRI Ch1, as this is the initial template for each model's morphology. These corrections range from 0.82 to 1.27 ($\sim$2-27\% correction). M81, M87, NGC~4395, Sombrero require only 2-9\% corrections, while NGC~4258, NGC~1052, and M94 have offsets of 14-27\%. We note that NIRSpec--MIRI flux offsets also exist in aperture extracted spectra of our targets and are even larger in some cases (2-32\% offsets). An example of our 1D spectra results are presented in Table~\ref{tab:1dspec}.

\begin{table}
	\centering
	\caption{\centering Decoupled 1d Spectra}
    \label{tab:1dspec}
    \begin{tabular}{ccccccc}
    \hline\hline
    \textbf{name} & 
    \textbf{wave} & 
    \textbf{agn\_flux}& 
    \textbf{gal\_flux} & 
    \textbf{mad} & 
    \textbf{aper\_flux} \\
    & ($\mathrm{\mu m}$) & ($\mathrm{mJy}$)  &  ($\mathrm{mJy}$) &  & ($\mathrm{mJy}$)
    \\
    \hline
        m87 & 1.65312 & 1.1420 & 31.5127 & 0.074 & 0.9695 \\
        m87 & 1.65351 & 1.1916 & 31.6278 & 0.067 & 0.9134 \\
        m87 & 1.65391 & 1.2060 & 31.7228 & 0.062 & 0.9096 \\
        m87 & 1.65430 & 1.2407 & 31.7783 & 0.055 & 0.9171 \\
        m87 & 1.65469 & 1.2401 & 31.8459 & 0.065 & 0.9173 \\
        m87 & 1.65509 & 1.2479 & 32.0359 & 0.070 & 0.9190 \\
        m87 & 1.65548 & 1.2621 & 32.1146 & 0.077 & 0.9272 \\
        m87 & 1.65588 & 1.2772 & 32.2810 & 0.079 & 0.9346 \\
        m87 & 1.65627 & 1.2863 & 32.2438 & 0.091 & 0.9403 \\
        m87 & 1.65667 & 1.2825 & 32.4211 & 0.052 & 0.9358 \\
        ... & ... & ... & ... & ... & ... \\
        \hline\hline
    \end{tabular}
    \\
    \vspace{1mm}
    \raggedright
      {\small \textbf{Note:} This table is available in its entirety in machine-readable form in the online journal. A portion is shown here for guidance regarding its form and content. We also include column 'pah\_f' to flag wavelengths in M94 that require a second fit template, see \S\ref{subsec:specialgalaxies}.}
\end{table}

\section{Results: ReveaLLAGN Nuclear Spectra Sample} \label{sec:results}

With continuous $1.6-24~\mu m$ combined 1D spectra for all seven ReveaLLAGN targets, we are able to analyse the resulting continua and compare results between nuclear extraction methods. We first present our decomposed AGN spectra in \S\ref{subsec:agnmodelspec}. We are then able to compare our decomposed AGN results to those from background subtracted aperture extracted spectra described in \S\ref{subsec:specextract} to highlight the reduced contamination of the AGN component that results from our iterative modeling method. We also present host galaxy flux contribution measured from our modeled galaxy components. Finally, we describe the methods applied to measure unresolved silicate emission features in our decomposed LLAGN spectra.

\subsection{Nuclear spectra}\label{subsec:agnmodelspec}

\begin{figure*}
    \centering
    \includegraphics[width=\linewidth]{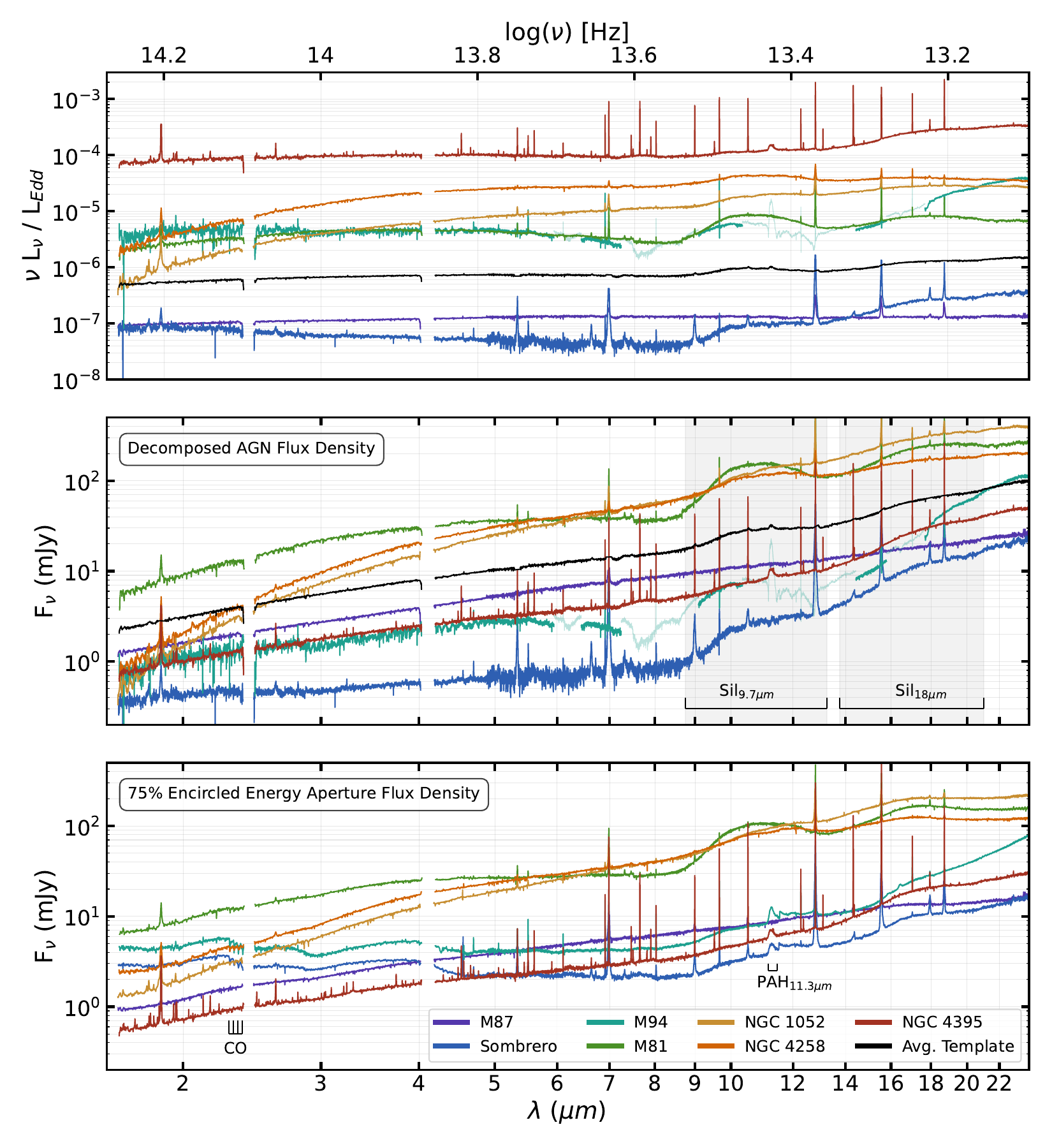}
    \caption{Nuclear Spectra from the ReveaLLAGN targets, comparing the results of our decomposed point-source AGN against those from a background-subtracted nuclear aperture.
    \emph{Top} -- Eddington-normalized luminosity of the unresolved AGN components, resulting from fitting and extraction methods described in \S\ref{sec:fit}.
    \emph{Middle} -- Unresolved AGN component flux density within 1.7'' aperture.
    \emph{Bottom} -- Nuclear spectra extracted using the methods of \citet{goold2026}. Despite background subtraction and aperture correction, contamination from surrounding host-galaxy emission is still apparent throughout the bottom panel, most notably at the CO bandhead absorption at $\rm \sim2.3~\mu m$ visible in Sombrero, M94, NGC4258, and M81 and also shortwards of 5~$\mu$m where the slope flattens relative to our decomposed AGN spectra, especially in M94 and Sombrero. Contamination from the nuclear host-galaxy flux in aperture-extracted spectra shows the importance of decoupling the AGN and galaxy components to extract accurate LLAGN continua. We include an average LLAGN IR spectrum template in black, made using our decomposed AGN spectra. Visible jumps in flux throughout the spectrum of M94 result from the separate models used to account for extreme PAH emission, represented in the top two panels by a more transparent line and} described in \S\ref{subsec:specialgalaxies}.
    \label{fig:spectra}
\end{figure*}

We present the decomposed AGN spectra of all seven ReveaLLAGN targets in Fig.~\ref{fig:spectra}. The top panel shows Eddington-normalized luminosities, while the second panel shows flux densities. M87 shows a nearly perfectly straight line across the full wavelength range, while M94 shows the most complicated spectrum due to some residual host galaxy contamination. Clear silicate emission features are seen between $\sim$$8-13\mu m$ in the spectra of all galaxies except M87 and NGC~4395, strongly suggesting that the silicate emission in these AGN, as characterized in \citet{goold2026}, is unresolved and on parsec scales. In \S\ref{subsec:silicate}, we investigate and quantify the compact nature of these features. Apart from the silicate emission, the spectra show consistent power-law slopes (i.e. are well described by straight lines) between 2.5 and 9~$\mu$m, with steepening at both shorter wavelengths and longer wavelengths. In the NIR, shortwards of 2.5~$\mu$m, NGC~1052 and NGC~4258 show a significant steepening. We explore the slopes and potential sources of steepening further in \S\ref{subsec:slopes}.

We note that jumps in the continuum of M94 result from the different models required to model its surrounding galaxy emission, described in \S\ref{subsec:specialgalaxies}. Despite these issues, its decomposed AGN spectra reveal a clear power-law shape in the near-infrared. In the PAH emission regions in M94, the difference in morphology of the PAH emission leads to significant uncertainty in the decomposed AGN spectra.

Therefore, in the top and middle panels of Fig.~\ref{fig:spectra}, we distinguish ranges of M94's decomposed AGN spectra which used the PAH fitting template with lighter line weight and increased transparency. We similarly use transparent markers throughout the paper to represent M94 values measured using data within these ranges.

We compose a LLAGN IR SED template from our decomposed AGN spectra, represented by the black spectrum in the top and middle panels of Fig.~\ref{fig:spectra}. We create this template using a weighted average of our spectra normalized between 5--6~$\rm\mu m$ -- excluding M94 due to the MIR fitting issues described above. We weight each spectrum using uncertainties characterized by the noise of its normalized flux density. Finally, we interpolate across emission features found in our targets to ensure a clean average continuum. Interpolation is applied across all lines measured by \citet{goold2026} with a window of 200~km~s$^{-1}$ (1500~km~s$^{-1}$ for strong Sombrero features), centered at the provided rest-frame wavelength. We include the averaged LLAGN SED template in Table~\ref{tab:1dspec} as a resource for future studies by the broader community.

We note that because our methodology focuses on accurate galaxy continuum subtraction, and emission lines have distinct morphologies, emission line features in our decomposed AGN spectra should be treated with caution. A full analysis of the nuclear emission lines is presented in \citet{goold2026}.

\subsection{Comparison to Aperture Method}\label{subsec:comparespec}

Further evidence of the success of our AGN isolation method is shown by comparing the point-source AGN spectra to those from background-subtracted aperture extraction methods (Fig. \ref{fig:spectra}, middle vs bottom). The $\rm 2.3~\mu m$ CO absorption features from galaxy contamination are clearly visible in three of the background-subtracted aperture spectra (M94, Sombrero, and NGC~4258); these are fully removed or minimized in our decomposed AGN spectra. Our AGN spectra also remove the flattening spectral slope in the NIR regions of any decomposed AGN spectra, which is also a signature of stellar flux contamination in the nuclear aperture spectra.

To test that our methods successfully removed host galaxy contamination, we measure the integrated flux across the largest CO feature at $\rm 2.29~\mu m$ before and after decomposition. The systems in which CO absorption features have the largest equivalent width (Sombrero and M94) in their aperture-extracted spectra measure only $\sim$3\% of the original flux after decomposition. Sombrero's decomposed AGN spectrum shows very weak reflections of the CO features in emission, but with a double-peaked shape indicative of potential kinematic differences between the galaxy and central region. Systems with low EW features in their aperture-extracted spectra measure between 5-36\% of the original after decomposition, but these values are representative of noise and remaining wiggles.

As is expected, M94 and Sombrero show the greatest changes, due to the especially low luminosity of their central engines and strong flux of their surrounding host galaxies. Sombrero's decomposed AGN spectra removes the stellar features seen in the aperture spectrum in the NIR, and the resulting NIR slope shows a power-law continuum.

While already visible in the aperture spectra, the silicate emission features at 9.7 and 18~$\mu$m are more prominent after decontamination. Our decomposed M94 AGN spectrum includes prominent silicate emission features previously hidden by galaxy emission in the nuclear aperture spectrum. Both methods show a steep slope in M94's spectra at wavelengths $\gtrsim17~\mu m$, extending beyond the 18~$\mu m$ silicate bump. We explore the strength of this emission further in \S\ref{subsec:silicate}.

\subsection{Galaxy Spectra}\label{subsec:galmodelspec}

\begin{figure}
    \centering
    \includegraphics[width=\linewidth]{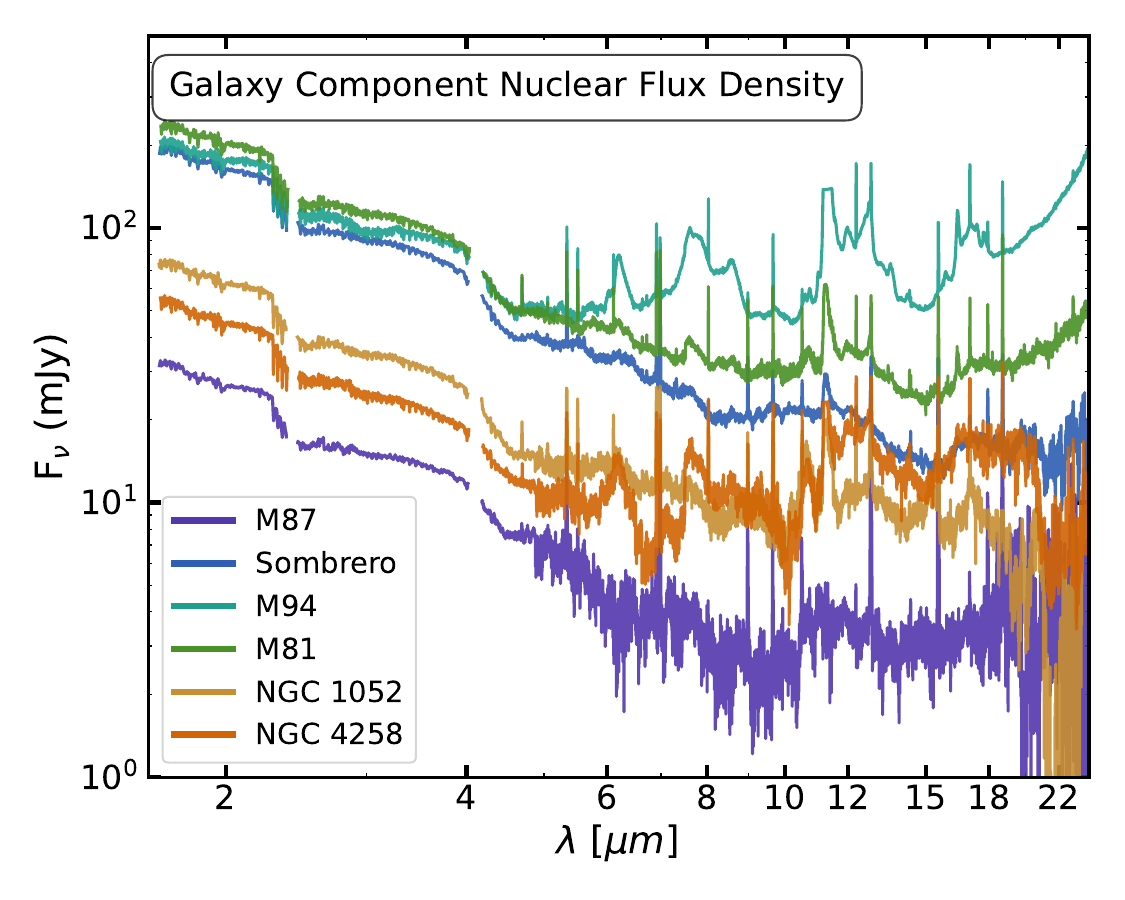}
    \caption{Host galaxy component spectra in the 1.7$''$ nuclear region. CO bandheads are clearly visible for all included targets at $\sim$2.3~$\mu$m. PAH features are well modeled for multiple galaxies, most notably M94, which necessitate special fitting methods as described in \S\ref{subsec:specialgalaxies}. We exclude NGC~4395 from our figure because its simplified galaxy emission requires fewer functions and is nonexistent in some channels.}
    \label{fig:galspec}
\end{figure}

Our galaxy component spectrum uses the same fixed $1.7''$ aperture as our AGN component. It is therefore not meant to directly correspond to the fraction of stellar contamination visible in aperture-extracted nuclear spectra. Rather, this flux corresponds to the overall continuum shape and prominent features of each host galaxy's contribution. Given the high signal-to-noise and the lack of existing high-quality nearby galaxy spectra between 2.5 and 5~$\mu$m, we include these spectra in Table~\ref{tab:1dspec} to make them available to the community. We include decomposed host galaxy spectra for our targets in Fig.~\ref{fig:galspec}, with the exception of NGC~4395, described in more detail at the end of this section. All six galaxy spectra in Fig.~\ref{fig:galspec} show similar NIR spectral shapes and absorption features, including clear CO absorption features at 2.3~$\mu$m. The MIR spectra are much more varied in their spectral shape, primarily reflecting differences in PAH emission. We observe distinct PAH emission features in all targets included in Fig.~\ref{fig:galspec}, except for M87, most notably at 11.3~$\mu$m.

The decomposed host galaxy fluxes of NGC~1052 and NGC~4258 share similarities with the spectra of the other targets at NIR wavelengths, indicating that the changing slope of their corresponding decomposed AGN fluxes in these wavelengths is not due to inappropriately weighted fitting function intensities in these channels.

NGC~4395 data processed through the \textit{JWST} pipeline resulted in strong negative backgrounds for most of its MIRI data. As such, NGC~4395 galaxy models include only a flat-sky function for MIRI channels, with the addition of a dim, diffuse \sersic{} function for NIRSpec data. All channel fits are overwhelmingly point-source dominated. We choose not to include this target's modeled galaxy flux in Fig.~\ref{fig:galspec} as its flux contribution is insignificant or nonexistent in the majority of the IR range.

We investigate the spatial origin of PAH emission in our decomposed AGN spectra by measuring the prominent 11.3~$\mu$m feature. We find no significant 11.3~$\mu$m emission in the aperture-extracted or decomposed AGN spectra of M87, M81, NGC~1052, or NGC~4258. Significant 11.3~$\mu$m emission exists in the aperture-extracted spectrum of Sombrero, but only $\sim10$\% of the feature's integrated flux remains after decomposition, with uncertainties large enough to indicate that the measurement likely reflects noise. The two targets with measurable 11.3~$\mu$m PAH emission in their decomposed AGN spectra, M94 and NGC~4395, both have important caveats: M94 has PAH morphology-related fitting issues, while the NGC~4395 result is affected by the background and simplified galaxy modeling described above. Combined with the clear PAH emission present throughout the host-galaxy spectra in Fig.~\ref{fig:galspec}, these results indicate that PAH emission in the decomposed spectra likely does not originate from an unresolved AGN component, but instead reflects residual host-galaxy emission.

\subsection{Comparison to Previous Isolation Method}\label{subsec:omairacompare}

We briefly compare our decomposed AGN spectra with decomposed PSF spectra from \citet{gonzolez-martin2025}. They decompose nuclear emission in 21 nearby AGN observations with MIRI/MRS, including 5/7 of the ReveaLLAGN targets. Their methods use a rotated synthetic WebbPSF with varying point-source amplitude and model the emission with two components: nuclear and extended emission, both approximated by 2D Gaussian profiles. We directly compare their resulting decomposed point-source spectra of NGC~1052 and M81 between 5--24~$\mu$m (shared via direct communication).

At wavelengths below 10~$\mu$m, our spectral fluxes match to within 5\%. At longer wavelengths, their spectra diverge gradually, with our spectra showing $\sim$15\% lower flux at the reddest wavelengths. Their spectra show higher fluxes in the MIR, resulting in even steeper MIR slopes than those we measure, as discussed further in \S\ref{subsec:slopes}. While we utilize a fixed 1.7$''$ aperture for all flux extraction and make no additional aperture correction, \citet{gonzolez-martin2025} extract their PSF spectrum using an aperture that changes with wavelength from 0.4$''$ at 5$\mu$m to 1.6$''$ at 27~$\mu$m and apply aperture corrections to account for flux losses. At 20~$\mu$m, the encircled energy at 1.7$''$ of our unresolved AGN component is roughly 95\%, thus it cannot account for the full difference we see between decomposed point-source fluxes. The difference likely comes from changes in the \textit{JWST} pipeline and flux calibration, since these are brighter targets almost completely dominated by AGN emission in the MIR.

\subsection{Measuring Silicate Emission}
\label{subsec:sil_measurement}

\citet{goold2026} measured 9.7~$\mu$m silicate strength from the 75\% encircled energy aperture-extracted spectra and found significant emission in 5/7 ReveaLLAGN targets. They directly compare nuclear aperture spectra of NGC~1052, NGC~4258, and M81 to their corresponding annular spectra, showing that these features are weak or absent at larger radii. This qualitatively demonstrates that existing silicate emission is strongest in nuclear regions of our observations. However, the presence of host-galaxy contamination in aperture-extracted spectra leaves the origin of these features uncertain. Using our decomposed AGN spectra, we can quantify the compactness of silicate emission in our LLAGN. To ensure that small jumps in flux at channel boundaries do not influence our measurements, we additionally stitched our decomposed AGN and aperture-extracted spectra at all MIRI channel boundaries, as well as boundaries of Ch2 medium and long subchannels, which occur in this wavelength range and contain the most prominent subchannel flux discrepancy. We also use decomposed AGN spectra measured with a constant 1.2$''$ aperture, since \citet{goold2026} calculates the 75\% encircled energy aperture assuming a full encircled energy of 1.2$''$. If the silicate emission in our targets is consistent with a completely unresolved component in our fits, the AGN-to-aperture flux ratio will be $\sim$1.33.

The continuum underlying each silicate feature was fit independently with a power law,
\begin{equation}
f_{\rm c}(\lambda)=A\lambda^{-\alpha},
\end{equation}
using a linear least-squares fit in log--log space. Before fitting the continuum, narrow emission lines were masked using the MIRI line lists from \citet{goold2026}. For each possible line, we excluded a velocity interval of $\pm 2000~{\rm km~s^{-1}}$, increasing this to $\pm 3000~{\rm km~s^{-1}}$ for Sombrero and M87, where broader features required a larger masking window. We also masked the 11.1--11.7~$\mu$m region to avoid contamination from the 11.3~$\mu$m PAH feature. For the 9.7~$\mu$m feature, the continuum was fit using the wavelength ranges 8.25--8.75 and 13.25--13.75~$\mu$m, and the silicate feature was measured over 8.75--13.75~$\mu$m. For the 18~$\mu$m feature, the continuum fitting windows were 13.25--13.75 and 21.0--22.0~$\mu$m, and the silicate feature was measured over 13.25--21.0~$\mu$m.

The integrated silicate flux was calculated from the continuum-subtracted spectrum, with masked regions linearly interpolated prior to integration.
\begin{equation}\label{eq:silicate_flux}
F_{\rm sil}
=
\int_{\lambda_1}^{\lambda_2}
\left[f(\lambda)-f_{\rm c}(\lambda)\right]d\lambda 
\end{equation}

where $\lambda_1$ and $\lambda_2$ are the integration limits for the feature. We measure an AGN-to-aperture silicate flux ratio, $F_{\rm AGN}/F_{\rm aper}$, where both fluxes are calculated using Equation~\ref{eq:silicate_flux} on the decomposed AGN spectrum and the 75\% encircled-energy aperture spectrum, respectively. This ratio acts as a diagnostic of how well the aperture-extracted feature is represented as an unresolved AGN component. Because the aperture spectrum contains 75\% of the flux from a point source, a value of $1/0.75\sim 1.33$ is the expected value for an ideal unresolved feature, but is not required for emission to be unresolved. $F_{\rm AGN}/F_{\rm aper}$>1.0 indicates that the aperture-extracted feature is well described as originating from a compact, parsec-scale component.

We also measured the equivalent width,
\begin{equation}
{\rm EW}_{\rm sil}
=
\int_{\lambda_1}^{\lambda_2}
\left[
\frac{f(\lambda)}{f_{\rm c}(\lambda)}
-1
\right]d\lambda .
\end{equation}
To characterize the width of the feature, we computed $W_{80}$, defined as the wavelength interval containing the central 80\% of the positive continuum-subtracted emission profile. To identify the peak wavelength, we first smoothed the continuum-subtracted residual spectrum using a running median with a box size of 51 spectral elements. We then searched for the maximum residual within 9.5--12.5~$\mu$m for the 9.7~$\mu$m feature and 15.5--20.5~$\mu$m for the 18~$\mu$m feature. 

The silicate strength, $S_{\rm sil}$, was defined as the logarithm of the ratio between the line-masked, interpolated spectrum and the fitted continuum at the peak wavelength. To reduce sensitivity to individual spectral elements, we calculated this ratio using the median interpolated flux density and median continuum level within a window centred on the peak wavelength, extending to the nearest 50 finite spectral elements on either side.
\begin{equation}
S_{\rm sil}
=
\ln
\left[
\frac{f(\lambda_{\rm peak})}
{f_{\rm c}(\lambda_{\rm peak})}
\right].
\end{equation}
With these definitions, positive values of both ${\rm EW}_{\rm sil}$ and $S_{\rm sil}$ correspond to silicate emission, while negative values correspond to absorption.

We present our silicate measurements for the decomposed AGN spectra in Table~\ref{tab:silicate}. Additional quantities are reported only for silicate features with measured strengths greater than 0.1. Consequently, NGC~4395 does not exhibit significant 9.7~$\mu$m silicate emission, while M87 shows no detectable silicate emission. We also exclude silicate measurements for M94 because noise and reduced flux reliability across its MIR spectrum prevent robust characterization of these features (see \S\ref{subsec:specialgalaxies}). The persistence and strength of silicate emission in the decomposed AGN spectra strongly supports the conclusion of \citet{goold2026} that the observed silicate features originate on parsec scales.

\begin{table*}
	\centering
	\caption{\centering Silicate Emission Measurements}
	\label{tab:silicate}
	\begin{tabular}{lcccccccccc}

        \hline \hline
        & \multicolumn{5}{c}{\textbf{9.7$\mu$m Silicate Feature}} & \multicolumn{5}{c}{\textbf{18$\mu$m Silicate Feature}}\\

        \cmidrule(lr){2-6}\cmidrule(lr){7-11}
		\textbf{Galaxy} &
        \textbf{S$_\text{Sil}$} &
        \textbf{F$_\text{AGN}$ / F$_\text{aper}$} &
        \textbf{Peak $\lambda$} &
        \textbf{EW} & 
        \textbf{W$_\text{80}$} &

        \textbf{S$_\text{Sil}$} &
        \textbf{F$_\text{AGN}$ / F$_\text{aper}$} &
        \textbf{Peak $\lambda$} &
        \textbf{EW} & 
        \textbf{W$_\text{80}$}
        \\
         & & & ($\mu m$) & ($\mu m$) & ($\mu m$) & 
          & & ($\mu m$) &($\mu m$) & ($\mu m$)
         \\
		\hline
        
            M87   & 0.013 & -- & -- & -- & -- & 
                    0.024 & -- & -- & -- & --\\

            Sombrero & 0.325 & 1.07 & 10.937 & 0.887 & 2.473 & 
                       0.271 & 1.58 & 17.321 & 1.183 & 3.722\\
                       
            M94   & -- & -- & -- & -- & -- & 
                    -- & -- & -- & -- & --\\            
            M81   & 0.778 & 1.28 & 10.609 & 2.766 & 2.324 & 
                    0.323 & 1.28 & 16.886 & 1.618 & 4.061\\            
            NGC~1052 & 0.192 & 1.24 & 11.011 & 0.455 & 2.173 &
                       0.162 & 0.90 & 16.431 &  0.756 & 4.212\\
            
            NGC~4258 & 0.264 & 1.14 & 10.418 & 0.759 & 2.392 &
                       0.108 & 0.68 & 16.918 & 0.387 & 3.434\\
            
            NGC~4395 & 0.046 & -- & -- & -- & -- & 
                       0.229 & 1.05 & 16.901 & 0.926 & 3.548\\
            
		\hline\hline
	\end{tabular}
    \\
    \vspace{1mm}
    \raggedright
      {\small \textbf{Note:} $\rm F_{AGN}/F_{aper}$ is the ratio of integrated flux measured from decomposed AGN spectra over that from the aperture-extracted spectra, as described in \S\ref{subsec:sil_measurement}. Features with $\rm S_{Sil}< 0.1$ do not include additional measurements. We choose to exclude M94 measurements, due to overlapping ranges of less-reliable flux values described in \S\ref{subsec:specialgalaxies}.}
\end{table*}

\section{Infrared--X-ray Flux Comparison} \label{sec:irx}

In higher-accreting AGN models, X-ray emission primarily originates from the reprocessing of UV emission through inverse Compton upscattering in a hot corona above the accretion disc. Another part of the UV emission is absorbed by surrounding dust and re-emitted in the IR as a "thermal" component, roughly consistent with a blackbody. Since most X-ray and IR emission in AGN share a common origin as reprocessed UV light, a correlation between the two is expected. Many previous comparisons have observed and established IR--X-ray relations for MIR \citep[e.g.][]{spinoglio1989,asmus2015, spinoglio2024} and NIR \citep[e.g.][]{alonso-herrero2001,burtscher2015,muller-sanchez2018} emission. These relations both tell us about the physics of the AGN, and can be used to estimate AGN bolometric luminosities and absorption.

However, most existing publications utilize comparison samples with log($\rm L_{Bol}/L_{Edd}$)$>$-3 and their interpretations often assume a unified model AGN. LLAGN SEDs lack a UV bump from a traditional thin disc and models suggest alternative emission mechanisms like RIAFs and truncated discs. Reprocessing by surrounding dust may also differ for LLAGN, as tori are expected to change, dissipate, or disappear completely at these low accretion rates. We compare relations between X-ray luminosities and IR emission from our decomposed LLAGN against the higher-Eddington samples of previous studies.

\subsection{Potential NIR Excess for LLAGN}\label{subsec:2um_xr_edd}

Previous work on NIR $K$-band (2~$\mu$m) AGN emission using NIR integral field spectroscopy has decontaminated galaxy and nuclear starlight emission through spectroscopic fitting \citep{burtscher2015,muller-sanchez2018,dumont2020}; this differs from the morphological approach we've taken here. 
In higher luminosity galaxies with $\rm L_{X} > 10^{41}~{ ergs~s}^{-1}$, 
\citet{muller-sanchez2018} and \citet{burtscher2015} found $L\rm _{NIR}$ vs. L$_{\rm X}$ relations with a nearly linear slope (between 0.85-0.98, depending on type), and interpreted the emission as coming most likely from hot dust. They also both find a clear offset in the normalization between Seyfert 1 and 2 galaxies, with the Seyfert 2s having lower NIR fluxes than Seyfert 1s, likely due to dust extinction. \cite{dumont2020} included a sample of 15 LLAGN covering a similar range of Eddington ratios as the ReveaLLAGN sample and found a much shallower $\rm L_{NIR}$ vs. $\rm L_X$ slope, with LLAGN having higher NIR flux at a given X-ray flux than expected from the higher luminosity relations. If interpreted as hot dust emission, the peak of this dust emission is expected at longer wavelengths; with our ReveaLLAGN targets, we can examine these trends and have continuous wavelength data out to longer wavelengths.

\begin{figure}
    \includegraphics[width=\linewidth]{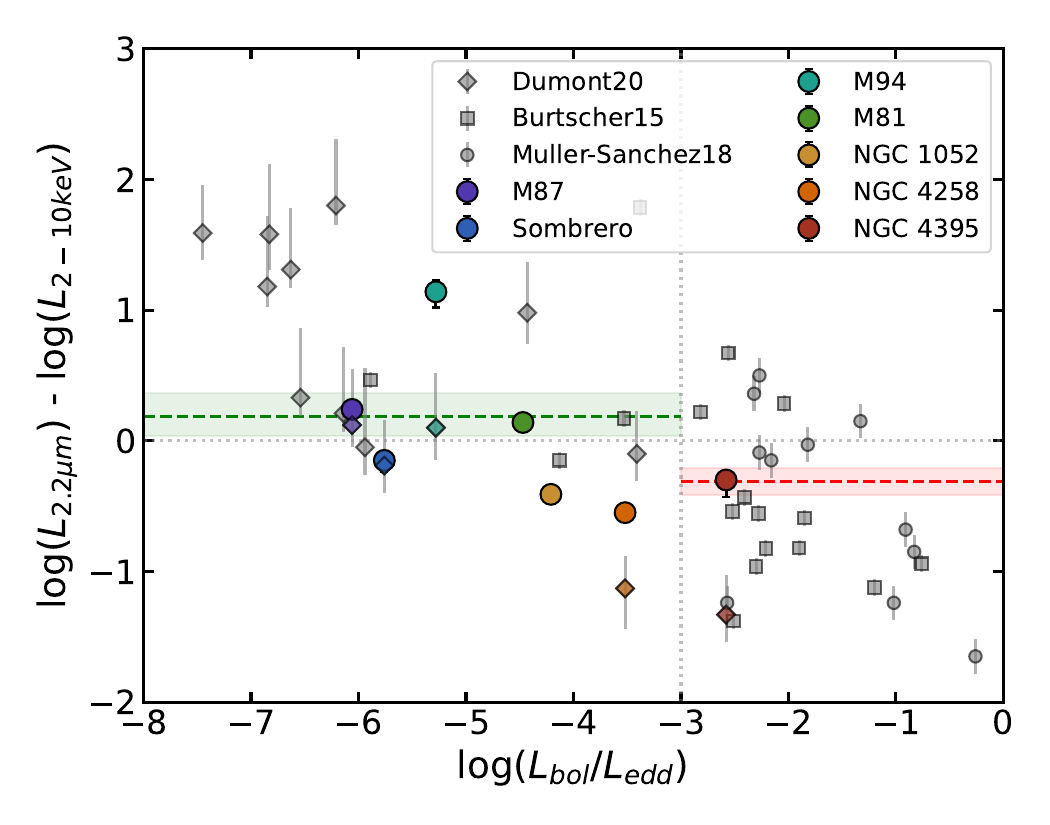}
    \caption{Ratio of integrated 2.2~$\mu$m and 2–10 keV X-ray luminosities vs Eddington ratio. Small points show data from \citet{dumont2020} (diamonds), \citet{burtscher2015} (squares), and \citet{muller-sanchez2018} (circles). Eddington ratios for comparison targets are calculated using black hole masses from \citet{bi2020}. The dashed horizontal green line shows the mean flux ratio for all data points where log($\rm L_{Bol}/L_{Edd}$)$\leq$-3, and the red line for those $>$-3. Shaded regions show $1\sigma$ errors calculated with bootstrapping. Our measurements support previous results suggesting a NIR flux excess for LLAGN.}
    \label{fig:nirxr}
\end{figure}

We integrate our seven decomposed AGN spectra from 1.99--2.43~$\mu$m to calculate K-band (2.2~$\mu$m) luminosities comparable to those presented in earlier works, applying calculated fractional errors described in \S\ref{subsec:specextract}. Fig.~\ref{fig:nirxr} compares this luminosity to the 2-10 keV X-ray emission as a function of Eddington ratio. We include literature comparison samples, using black hole mass values from \cite{bi2020} and X-ray-based bolometric corrections from \citet{duras2020} to calculate Eddington ratios. The AGN samples of \cite{burtscher2015} and \citet{muller-sanchez2018} primarily use 14-195 keV X-ray luminosities from SWIFT/BAT \citep{baumgartner2013} to minimize the effect of obscuration, and both apply 2-10 keV archival Chandra luminosities to a small number of sample AGN. \citet{muller-sanchez2018} directly provides $\rm log(L_{2-10keV})$ for their sample, converting between hard and soft X-ray luminosities using $\rm log(L_{2-10kev})=1.06*log(L_{14-195kev}) - 3.08$ \citep{winter2009}. For consistency, we use this same relation to convert all \citet{burtscher2015} luminosities to $\rm log(L_{2-10kev})$. We apply provided median $\rm L_X$ and $\rm L_{2\mu m}$ errors from each sample for plotting and weighted analysis.

For shared targets in the \citet{dumont2020} sample, we apply our M$_{\rm BH}$ and log(L$_{\rm X}$) values and plot them with matching color in Fig.~\ref{fig:nirxr}. We find NIR luminosities that are significantly higher than those in \citet{dumont2020} for M94, NGC~4258, and NGC~4395. Variability may partially contribute to these discrepancies, or they may be due to differences in method. The decomposition of AGN in \citet{dumont2020} was spectroscopic (through the reduction in strength of CO bandheads), while our method is morphological. Our method enables us to identify unresolved AGN emission, but it is also possible that some weak stellar emission is included. On the other hand, the spectroscopic fits yield large uncertainties in flux, as is apparent from the error bars in Fig.~\ref{fig:nirxr}.

Our sample shows consistency with these data from \citet{dumont2020} suggesting larger NIR-to-X-ray luminosities for low Eddington ratio systems relative to those seen at higher Eddington ratios. We note that our decomposition methodology results in much lower errors than the spectral decomposition method in \citet{dumont2020}, but we have fewer targets than that archival ground-based study.  
Dividing sources at log($\rm L_{Bol}/L_{Edd}$) of -3, for the low Eddington sources we find a weighted mean $\rm log(L_{2.2\mu m})-log( L_{2-10keV})$ of 0.203$\pm$0.15, where the 1$\sigma$ error is determined from bootstrapping. For log($\rm L_{Bol}/L_{Edd}$) > -3, we find the luminosity/flux ratio is significantly lower with $\rm log(L_{2.2\mu m})-log( L_{2-10keV})$ of -0.335$\pm$0.11. Thus, there is a $\sim$0.54 dex (factor of 5) enhancement in the NIR emission of low-Eddington AGN relative to higher-Eddington AGN. 
Repeating this process using unweighted mean increases the difference in ratios to roughly one magnitude; 0.432$\pm$0.15 for log($\rm L_{Bol}/L_{Edd})<-3$ and -0.369$\pm$0.10 for log($\rm L_{Bol}/L_{Edd}$)~$>-3$.

Our sample's full IR SEDs can potentially shed light on the emission mechanism of AGN in NIR. While the NIR light in luminous AGN is thought to be dominated by hot ($\sim$1000~K) dust, the lack of any thermal bumps in our SEDs suggests the NIR emission may instead include a compact-jet contribution, which could explain the relatively higher NIR-to-X-ray luminosity ratios.

LLAGN are often radio-loud \citep{ho2001,ho2002} and can host parsec-scale jets in their unresolved radio cores \citep{Nagar2005,mezcua2014,baldi2018}. To test whether the NIR excess in our LLAGN could arise from jet emission, we compare NIR-to-X-ray luminosity ratios with a nuclear radio-loudness indicator. Following the methods of \citet{FernandezOntiveros2021}, we define radio-loudness as the flux-density ratio $F_{\rm 8.4~GHz}/F_{\rm 2-10~keV}$ and assume this quantity tracks the strength of jet emission. We adopt this radio-to-X-ray ratio instead of the more classical radio-to-optical because the latter is typically dominated by host-galaxy emission in LLAGN. We use 8.4 GHz measurements to sample a high-frequency regime with broad coverage and restrict the analysis to Very Large Array (VLA) data, as its high angular resolution minimizes contamination from extended radio lobes. Archival radio flux densities were retrieved from NED and supplemented with additional measurements from \citet{thean2000}. For M94 and NGC~4395, where no direct measurements were available, we interpolate 8.4 GHz flux density from corresponding 5 and 15 GHz VLA measurements.

\begin{figure}
    \includegraphics[width=\linewidth]{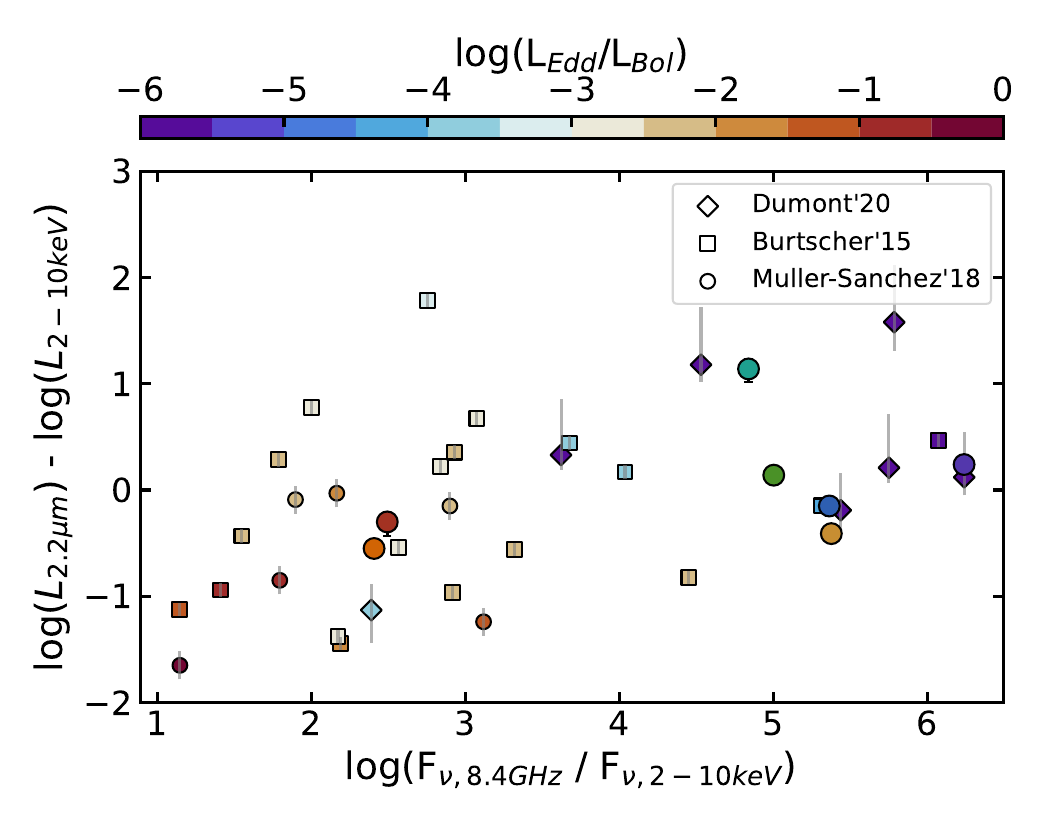}
    \caption{Ratio of integrated 2.2~$\mu$m and 2–10 keV X-ray luminosities vs Radio Loudness. Small points show data from \citet{dumont2020} (diamonds), \citet{burtscher2015} (squares), and \citet{muller-sanchez2018} (circles). Comparison data are colored by Eddington ratios, calculated using black hole masses from \citet{bi2020}. Our measured correlation support the jet origin for LLAGN NIR flux excess shown in \ref{fig:nirxr}.}
    \label{fig:nirrad}
\end{figure}

Fig.~\ref{fig:nirrad} compares radio-loudness with $\rm log(L_{2.2\mu m})-log( L_{2-10keV})$ for the majority of targets in Fig.~\ref{fig:nirxr}. Our results recover the established inverse correlation between log($\rm L_{Bol}/L_{Edd}$) and radio-loudness \citep{ho2001,fender2003}, with low-Eddington sources generally occupying the more radio-loud regime. We find a statistically significant positive correlation between the NIR excess and radio-loudness (Pearson's $R=0.445$ and $p=0.002$). Taken together, these correlations are consistent with a compact jet origin for the NIR excess measured in LLAGN. A dedicated analysis, including broader-wavelength spectral energy distribution modeling in future work (Prieto et al., {\em in prep}; Duncan et al., {\em in prep}), will be needed to confirm the origin of the NIR excess, quantify contribution from compact jets, and trace the strength of X-ray continua.

\subsection{12~$\mu$m--X-ray Luminosity Relation}\label{subsec:12um_xray}

When considering AGN emission in the MIR, 12~$\mu$m is commonly used. This overlaps with the 7-12~$\rm \mu m$ mid-IR "pivot" wavelength from \citet{spinoglio1989} where normalized luminous AGN SEDs show the least scatter, where absorption and reemission of the initial continuum are balanced. The IRAC $\rm 12~\mu m$ band has been used for AGN sample selection \citep[e.g.][]{rush1993, spinoglio1995}, and the correlation between $\rm 12~\mu m$ and X-ray emission has been measured \citep[e.g.][]{gandhi2009,ichikawa2019}.

\begin{figure}
    \includegraphics[width=\linewidth]{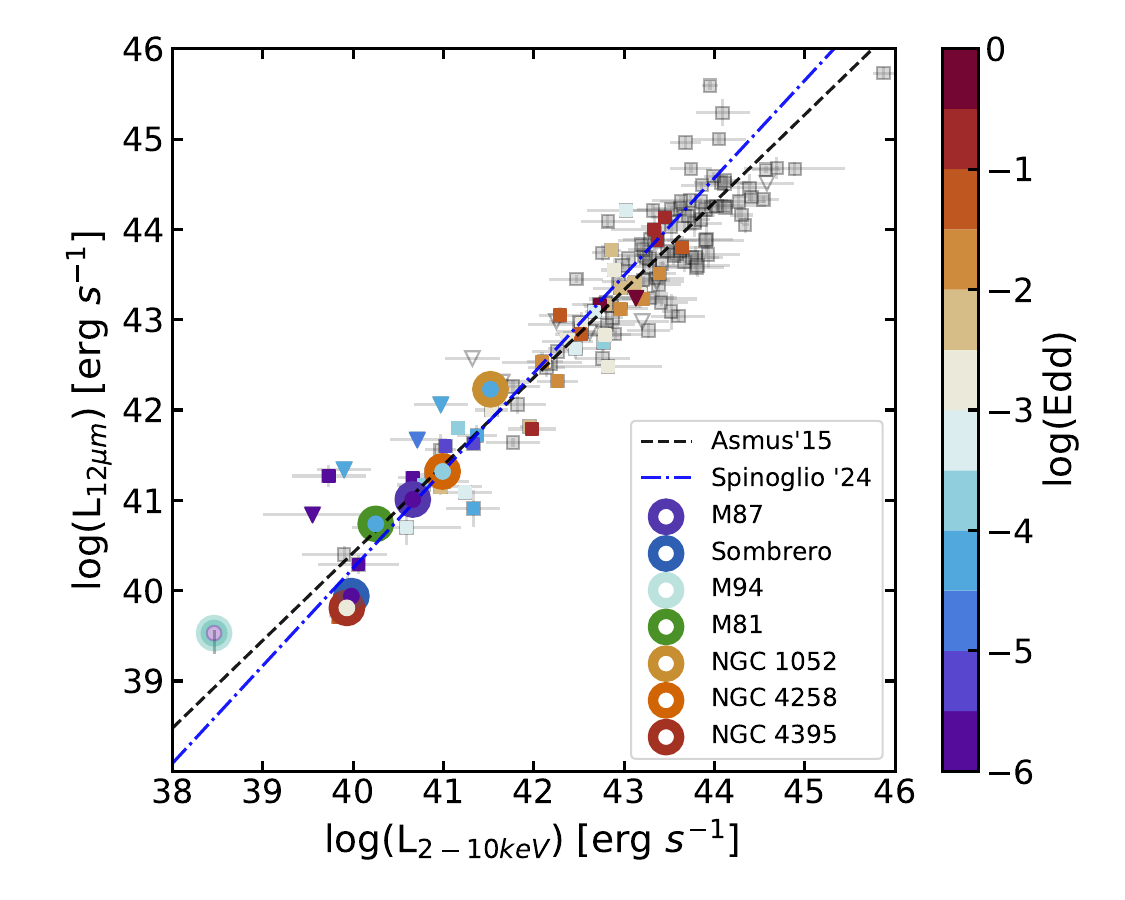}
    \caption{Relation of integrated 12~$\mu$m and 2–10 keV luminosities, compared to values (squares) and upper limits (triangles) from \citet{asmus2015}. Colormap shows Eddington ratio where available. We include linear fits to this relation from both \citet{asmus2015} and \citet{spinoglio2024}. All seven ReveaLLAGN targets align closely to previously established relations. We calculate $\rm log(L_{Bol}/L_{Edd})$ for the comparison sample with $\rm M_{BH}$ values from \citet{bi2020} and methods described in \ref{appendix:eddington}. We apply transparency to M94 due to its less reliable flux, see \S\ref{subsec:agnmodelspec}.}
    \label{fig:12xrratio}
\end{figure}

We integrate flux from 11.5--12.5~$\mu$m to calculate $\rm L_{12 \mu m}$ for our targets to test the correlation with 2-10 keV X-ray luminosity for decomposed point source AGN. Fig.~\ref{fig:12xrratio} shows this correlation, with comparison against the sample of \citet{asmus2015}. Our data probes low luminosities; extending 1.09 and 0.25 dex lower than the comparison sample in $\rm log(L_{2-10keV})$ and $\rm log(L_{12 \mu m})$ respectively.

We calculate $\rm log(L_{Bol}/L_{Edd})$ for comparison targets using black hole masses from \citet{bi2020} when available. We also include linear relations from both \citet{asmus2015} and \citet{spinoglio2024}, which have slopes of 0.98 and 1.08, respectively. Our target LLAGN align well with both linear relations, with scatter around the included relations similar to that of comparison data.

\citet{asmus2015} states that Radio-loud AGN (correlating with low Eddington rates) within their sample exhibit a higher MIR-X-ray ratio at low luminosities, citing the potential for significant jet contribution. To analyse this claim, we repeat the process described in \S\ref{subsec:2um_xr_edd} to calculate average $\rm log(L_{12 \mu m})-log(L_{2-10keV})$, using our seven targets and all comparison data with available Eddington ratio, again dividing sources at $\rm log(L_{ Bol}/L_{Edd})=-3$. The low-Eddington sample shows MIR emission with a weighted mean 0.13$\pm$0.10 dex higher than the high-Eddington sample; thus, we find no significant difference between the two.


\section{Discussion}\label{sec:discussion}

\subsection{Silicate Emission}\label{subsec:silicate}

Amorphous silicates in the circumnuclear dust surrounding AGN generate two broad features in their MIR spectra. Stretching of the Si-O bond produces a well-studied feature centred near 9.7~$\mu$m, while bending of the O-Si-O bond produces a feature centred near 18~$\mu$m. Depending on the geometry and optical depth of the dust distribution, these features can appear as either emission or absorption. Silicate absorption typically occurs in the average spectra of Type-2 AGN \citep{sturm2006,hernan-caballero2011}, in which the cool dust of the edge-on torus obscures the central engine. Silicate emission is commonly observed Type 1 AGN because their relatively unobscured line of sight provides a direct view of warm, illuminated silicate dust near the nucleus \citep{hao2005,wu2009}. However, emission features can appear in spectra of Type-2 AGN \citep{mason2009} and absorption features are present in spectra of some Type-1 AGN \citep{hatziminaoglou2015}. Both emission and absorption are observed in LINERs and LLAGN \citep{sturm2005,Mason2012,zhang2026}.

We describe our silicate measurements in \S\ref{subsec:sil_measurement} and  Table~\ref{tab:silicate} lists the results of silicate measurements of our decomposed LLAGN spectra.  We find five of our seven decomposed AGN spectra have clear silicate emission; none show absorption.

The silicate AGN-to-aperture flux ratio ($\rm F_{AGN}/F_{aper}$; see \S\ref{subsec:sil_measurement}) is a diagnostic of whether the aperture-extracted feature is consistent with unresolved, point-like emission. Ratios above 1.0 support a highly unresolved feature, values near 1.33 indicate an idealized point-like origin, and values above 1.33 suggest host galaxy contamination may have weakened the associated feature in an aperture extracted spectrum. The strength of the 9.7~$\mu$m features after decontamination and their corresponding high $\rm F_{AGN}/F_{aper}$ values strongly supports the conclusion of \citet{goold2026} that silicate emission in our targets is not due to surrounding host-galaxy contamination, and likely originates from compact circumnuclear dust at parsec spatial scales. Taking the MIRI PSF FWHM of $\sim0.4''$\footnote{Obtained from \href{https://jwst-docs.stsci.edu/jwst-mid-infrared-instrument/miri-performance/miri-point-spread-functions}{JDox} as of 2026 June 5.}, which sets the approximate resolution limit at silicate wavelengths, unresolved emission corresponds to physical scales ranging from $\sim7.2$~pc (M81) to $\sim37.6$~pc (NGC~1052).

We note that the 9.7~$\mu$m AGN-to-aperture flux ratio of M81 indicates its silicate emission is almost completely unresolved, despite being the nearest target. Additionally, if we estimate sublimation radii using the methods of \citet{nenkova2008} and an assumed sublimation temperature of 1500~K \citep{xie2017}, the estimated $\rm R_{sub}$ values for all seven targets are smaller than the minimum spatial scales resolvable by MIRI at these wavelengths and therefore remain unresolved. 

Measured $\rm F_{AGN}/F_{aper}$ for 18~$\mu$m features vary more between targets relative to the 9.7~$\mu$m feature, with NGC~4258 showing a significantly decreased strength in its decoupled spectra. Flux from NGC~1052 and NGC~4395 remain relatively unchanged after decoupling. The 18~$\mu$m flux ratio of 1.58 from Sombrero implies that the surrounding galaxy was partly diluting or masking a genuinely compact nuclear dust component in its aperture-extracted spectrum. The difference in $\rm F_{AGN}/F_{aper}$ values between 9.7~$\mu$m and 18~$\mu$m features from the same AGN is reasonable, as the they originate from different populations of dust and either feature can be observed in emission, absorption, or missing in a single AGN spectrum \citep{hatziminaoglou2015}.

Galactic ISM silicates (sometimes called "astronomical" silicates) peak at 9.7~$\mu$m \citep{chiar2006,henning2010}. In our decomposed spectra, the 9.7~$\mu$m Si-O emission features peak at $\rm >10.4 \mu m$. Inversely, our measured 18~$\mu$m features peak at shorter values, $\rm <17.3~\mu m$. These are both consistent with silicate emission measured from IR spectra of higher-luminosity AGN samples \citep[e.g.][]{sturm2005,smith2010,martinez-paredes2020,reyes-amador2024}. In addition to peak wavelength, we include equivalent width and feature width ($\rm W_{80}$) measurements in Table~\ref{tab:silicate}, as these parameters can provide information about the chemical composition, size distribution, and other properties of silicate grains in our LLAGN \citep{hatziminaoglou2015,xie2017,martinez-paredes2020}.

Some models predict the disappearance of a classical dusty torus in LLAGN, whether due to a lack of mass inflow to the central engine, or because accreted mass outflows shift from disc winds to jets \citep{elitzur2006,vollmer2008}. The existence of silicate emission in our decomposed spectra argues that warm dust on parsec scales still clearly survives in many LLAGN. The presence of silicate emission in most targets could suggest diffuse and optically thin remnants of a dissipating torus \citep{mason2013}. We explored whether the NIR excess discussed in \ref{subsec:2um_xr_edd} correlates with $\rm S_{9.7}$ and find no correlation. This supports the idea that, while warm dust survives close to the central engine, the NIR excess likely traces an additional component e.g. jet emission.

Using \textit{JWST} data for five of the ReveaLLAGN targets, \citet{gonzolez-martin2025} fit a variety of models to their mid-infrared SEDs. Overall, they find that their models don't provide good fits to all of our sources, with three of the five having unsatisfactory SED fits. In three of these targets (Sombrero, M81, and M94), the best fit model is the   \citet{nenkova2008} clumpy torus model, which is not the best-fit model in any of the higher luminosity targets in their sample. In clumpy torus models, the dust is arranged in discrete clouds, which allows emission from warm illuminated surfaces to escape through gaps in the torus. As a result, weak or complex silicate emission features can be observed in these systems. However, the \citet{nenkova2008} clumpy models do not allow silicate emission strong enough to match that observed in some LLAGN \citep{smith2010,mason2013}.

Silicate emission may also originate from collimated dust filaments with parsec-scale widths that pass near the nucleus, which are seen as a primary cause of obscuration and a potential fueling channel for the AGN. A single filament passes immediately adjacent to the nucleus of NGC~1052 \citep{prieto2021}. In Sombrero, these filaments are optically thin near the nucleus but become increasingly optically thick beyond that scale \citep{prieto2014}. This scenario is compatible with the presence of strong unresolved silicate emission in the spectra of these targets. By comparison, M87 contains little to no central dust \citep{prieto2021}, consistent with its flat decomposed continuum and absence of silicate features.

\subsection{Comparison of IR Slopes to those of Higher-Luminosity AGN} \label{subsec:slopes}

We study the overall SED slopes of our decomposed AGN spectra, investigate the potential impacts of dust on these measurements, and compare the near- and mid-infrared slopes of our targets to those of higher-luminosity AGN. Our spectra (Fig.~\ref{fig:spectra}) show power-law slopes over large ranges in wavelength. To characterize this, we use \textsc{lmfit} to fit a power law function to our decomposed AGN spectra at NIR and MIR wavelength ranges. We mask known emission features using a consistent width of 1500 $\rm km~s^{-1}$ to fit only the continuum shape. For the NIR, we fit between 2.8--5.2~$\mu$m, chosen for its low noise and overlapping range with \textit{WISE} W1-W2 color. MIR slope is measured between 13--26$\mu$m, in which we also mask the region of strong overlapping silicate features present in the majority of our targets between 8.5--22~$\mu$m.

\begin{figure*}
    \centering
    \includegraphics[width=\linewidth]{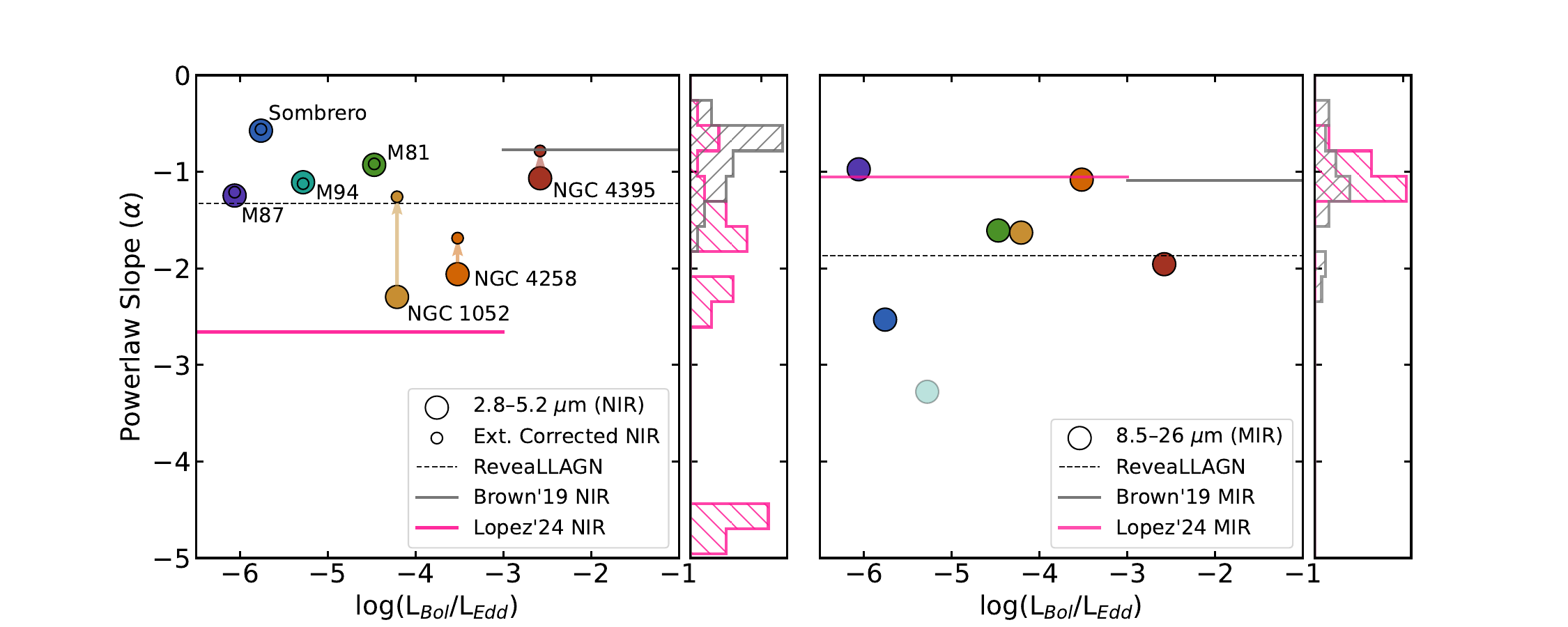}
    \caption{Power law slopes from fits to AGN continuum vs. Eddington ratio. Power laws were fit in both NIR (Left) and MIR (Right) wavelength ranges. Arrows and small circles indicate the shift in NIR slope after applying $\rm N_H$-based extinction corrections. Silicate features are masked for the measurement of MIR slopes. The black dashed line marks the average slopes of the ReveaLLAGN sample. We include slopes fit in matching wavelength ranges to 32 higher-luminosity AGN SEDs from \citet{brown2019} (grey), and to LLAGN model SEDs from \citet{lopez2024} (pink). Histograms show the distribution of comparison sample slopes and horizontal lines show their average slopes in corresponding colors. In the right figure, we apply transparency to M94 due to its less reliable flux throughout the MIR, see \S\ref{subsec:agnmodelspec}.}
    \label{fig:slopes}
\end{figure*}

We present the resulting slopes of these power-law fits in Fig.~\ref{fig:slopes}. The mean slopes from our seven LLAGN are -1.37 for NIR, and -1.99 for MIR. The formal errors on these slopes are extremely low (on the order of $\sim$0.001). There are systematic uncertainties in the MIR range caused by the varying presence and strength of silicate emission features between targets. We characterize this uncertainty by also fitting to masked regions surrounding the first feature (7--14$\mu$m) and second feature (13--26$\mu$m). The variation from this method gives a mean MIR slope uncertainty of 0.33, with our steepest measurement in M94 varying the most by 1.04.

{\bf Impacts of Dust on NIR slopes:} NGC~1052 and NGC~4258 are notable outliers in the power-law slope in the NIR, measured in Fig.~\ref{fig:slopes} and clearly visible in the decomposed spectra of Fig.~\ref{fig:spectra} as a steep drop in flux at the shortest wavelengths. We consider the possibility that this NIR flux drop is due to IR extinction, given that they both have the highest X-ray measured hydrogen column density, with both log($N_{\rm H}/{\rm cm}^{-2}) > 22$. While the exact source of extinction for the X-ray emission isn't well known, if we make the assumption that it is due to foreground extinction that also affects IR wavelengths, we can apply extinction correction derived from log($N_{\rm H}$). We use the relation $N_{\rm H}(\rm cm^{-2})=(2.21 \pm 0.09)\times10^{21} A_{V}$ \citep{guver2009}, and apply the resulting $A_{V}$ to a \citet{gordon2023} Milky Way (MW) extinction curve with $R_{V}=2.3$. Effects of this correction on all seven LLAGN are shown by arrows leading to small circles in Fig.~\ref{fig:slopes}. The $N_{\rm H}$-based correction results in a flattened power law continuum shape for both targets in the NIR, with measured power-law slopes of -1.31 for NGC~1052 and -1.73 for NGC~4258, making their slopes comparable to the other ReveaLLAGN targets.

We note that the X-ray and optical/infrared extinction are not necessarily expected to be similar, due to both to varying gas-to-dust ratios (with the X-ray absorption coming from any gas, while the infrared absorption is just due to dust) and because the X-ray and infrared may not probe the same sightlines.

{\bf Comparison to IR slopes of higher-Eddington AGN:} We repeat the same fitting process using 41 SEDs of higher-luminosity AGN from \cite{brown2019}. SEDs in this sample span 0.9-30~$\mu m$ by combining multiwavelength photometry and archival spectroscopy, using interpolation and extrapolation to fill gaps between spectral measurements. We mask interpolated and extrapolated flux data and exclude SEDs with fewer than 10 remaining data points within the chosen wavelength ranges. This leaves a comparison sample of 32 AGN, with a median NIR slope of -0.788$\pm$0.50 and MIR slope of -1.10$\pm$0.49. Fig.~\ref{fig:slopes} shows the distributions and averages of the \citet{brown2019} slope measurements in grey. We employ a two-sample Anderson-Darling test to quantify the similarity of slope distributions between our LLAGN and those of the higher-Eddington sample. We calculate p-values of 0.04 for NIR slopes and 0.02 for MIR slopes, indicating that the LLAGN and higher-luminosity AGN slope distributions differ below $\rm p=0.05$, but above the stricter $\rm p=0.01$ threshold. Given the potential influence of PAH emission on M94's exceptionally steep MIR slope, we repeat the Anderson-Darling test excluding this source and find weaker evidence for a difference between the MIR slope distributions ($\rm p=0.08$).

{\bf Physical Interpretation:} The two-sample Anderson-Darling test implies that the continuum slopes of our LLAGN are statistically distinct from the higher-Eddington AGN sample, particularly in the MIR where we see a wider range of slopes extending to steeper values. However, the NIR slope results are likely influenced by the outlier values from NGC~4258 and NGC~1052. If we assume that these steeper NIR slopes are due to dust extinction and repeat the Anderson-Darling test using extinction-corrected measurements described above, we obtain a p-value of 0.16. This result indicates that extinction-corrected NIR slopes are broadly similar to those of the \citet{brown2019} comparison sample. This would be consistent with previous work that found IR SED slopes of LLAGN within the range of typical Seyferts \citep{mason2013}. We note, however, that this does not imply that the physical mechanism driving the correlation is the same between LLAGN and higher-luminosity AGN. For example, in bright AGN, the balance between MIR and X-ray luminosities seems to be produced by UV photons from the accretion disc. LLAGN do not have such a source of UV photons. However, processes like inverse Compton scattering could maintain a balance between IR and X-ray fluxes, keeping the sources around a similar correlation, as we show for our targets in \S\ref{subsec:12um_xray}. \citet{FernandezOntiveros2023} show that their LLAGN similarly follow the same correlation; however, none of their SEDs show a thermal UV bump that could explain the dust heating. Our MIR slopes are more varied and, overall, steeper than those of the comparison sample, although the evidence for this difference is weaker when we exclude M94 because of its possible MIR PAH contamination. We will test our decomposed continuum spectra against detailed physical jet models in an upcoming paper (Duncan et al., {\em in prep}) where we show evidence for the need for cool disk components.

\subsection{Comparison to Model-Based LLAGN SEDs}\label{subsec:models}

We also use our decomposed AGN spectra to test existing SED models for LLAGN. The \citet{lopez2024} LLAGN \textsc{CIGALE} module applies templates from \citet{nemmen2014} and \citet{schartmann2005} to model LLAGN SEDs, assuming emission from combined ADAF and truncated disc. They test this module on a sample of 50 X-ray detected LINERs and Seyferts to fit their SEDs. We follow their methods and consider 46/50 galaxies with fits resulting in reduced $\chi^2<5$ for analysis. In their LLAGN models, they define the parameter $\rm \delta_{AGN}$ as the ratio between ADAF and classic standard-disc emission, with $\rm \delta_{AGN}=0.0$ corresponding to purely ADAF and $\rm \delta_{AGN}=1.0$ to purely disc. Their fitting process allowed three possible fixed $\rm \delta_{AGN}$ values that primarily affect the NIR power-law: 0.0, 0.25, or 0.5. In addition to the affects of $\rm \delta_{AGN}$ on their SED fits, we consider three important parameters they incorporate in the \textsc{SKIRTOR} AGN module \citep{stalevski2012,stalevski2016} to model AGN emission, where each parameter was allowed one of two values: inclination angle ($i$) of $10^{\circ}$ or $80^{\circ}$, torus opening angle ($OA$) of $10^{\circ}$ or $70^{\circ}$, and optical depth at $\rm 9.7~\mu m$ ($\tau_{9.7}$) of 3 or 11.

\begin{figure}
    \centering
    \includegraphics[width=\linewidth]{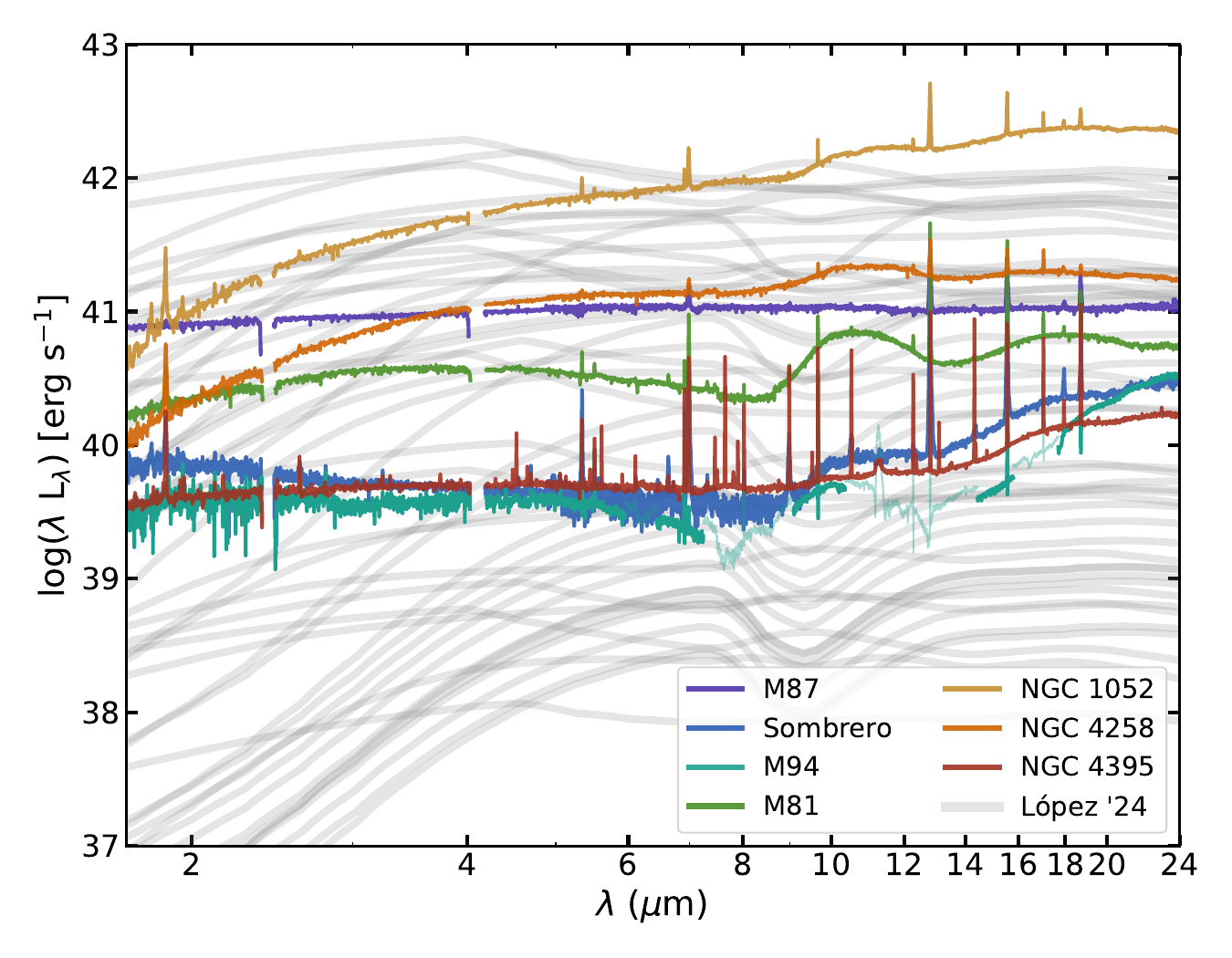}
    \caption{Decomposed AGN spectra of ReveaLLAGN targets, compared to 47 LLAGN model-fit SEDs from \citet{lopez2024} (grey lines). MIR slopes of our decomposed AGN spectra show inconsistency with a significant fraction of model SEDs, and our spectra have considerably steeper MIR slopes. While silicate features in our spectra are either absent or in emission, the model SEDs show silicate features primarily in absorption and with peak wavelengths shortward of our measurements.}
    \label{fig:modelcompare}
\end{figure}

We compare our decomposed AGN spectra with the model-fit LLAGN SEDs of \citet{lopez2024} in Fig.~\ref{fig:modelcompare} and find several differences. The model-fit SEDs have steeper NIR slopes than our spectra on average, while their MIR slopes are shallower than those we measure. While five of our seven targets show 9.7~$\mu$m silicate emission, the model-fit SEDs primarily show this feature in absorption. Even the SEDs with silicate emission peak at 9.7~$\mu$m, shortward of the peaks in our spectra at $>10.3~\mu$m (see \S~\ref{subsec:silicate}). Finally, the \textsc{CIGALE} module does not incorporate a jet component and therefore cannot reproduce the nearly straight, jet-dominated power-law spectrum of M87.

Fig.~\ref{fig:slopes} shows the mean slopes of power laws fit to the NIR and MIR ranges of the \citet{lopez2024} SEDs with pink horizontal lines, using the same methods described in \S~\ref{subsec:slopes}. Pink histograms show the distributions of the power-law slopes for these model SEDs. Their average NIR slope of -2.66 is steeper than the average slope of -1.37 measured for our spectra, while their average MIR slope of -1.05 is shallower than our average of -1.99. We note that their steeper average NIR slope is influenced by a distinct distribution of exceptionally steep values.

We repeat the two-sample Anderson-Darling test as in \S~\ref{subsec:slopes} to compare the continuum slopes of our decomposed AGN spectra to those of the \cite{lopez2024} model SEDs. If we consider the full distribution of NIR slopes, we calculate a p-value of 0.02, indicating that the two distributions are statistically distinct. This is the result of 16/46 of the fitted models having NIR slopes $<-4$; these steep sloped models all have significant AGN fraction with $\delta_{AGN}=0.5$. All 16 share the same \textsc{SKIRTOR} parameters of $i=80^{\circ}$, $OA=70^{\circ}$, and $\tau_{9.7}=11$. Beyond the effect of higher $\delta_{AGN}$ causing steeper power-law slope, these associated parameter values support extinction as a potential cause of steeper observed NIR slope discussed in \S~\ref{subsec:slopes}. However, the NIR slopes are far lower than even our steepest measured values in NGC~1052 and NGC~4258. In contrast, our observed MIR slopes of our targets are consistently steeper than those in the \cite{lopez2024} sample regardless of their associated \textsc{CIGALE} parameters, with an Anderson-Darling test returning a p-value of 0.003.

\begin{figure*}
    \centering
    \includegraphics[width=\linewidth]{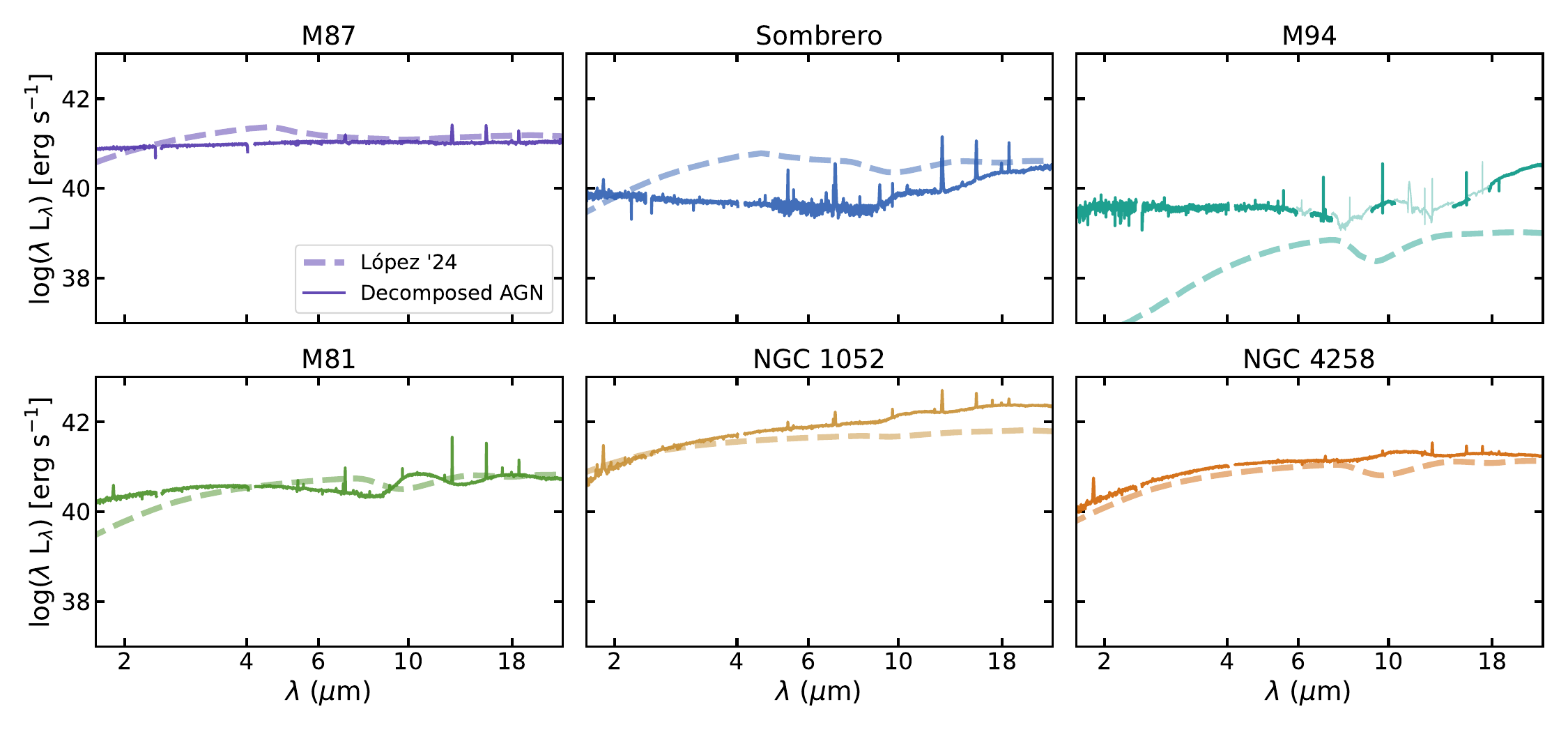}
    \caption{Direct comparison of the decomposed AGN spectra for six ReveaLLAGN targets (solid lines) with their corresponding model-fit SEDs from \citet{lopez2024} (dashed lines). The models reproduce the continuum shapes of M81, NGC~1052, and NGC~4258 more closely than those of M87, Sombrero, and M94, but do not reproduce the observed silicate emission. The M87 model also fails to reproduce its jet-dominated power-law spectrum.}
    \label{fig:sedspeccompare}
\end{figure*}

We present direct comparisons between six of our decomposed AGN spectra and their corresponding model-based SEDs from \citet{lopez2024} in Fig.~\ref{fig:sedspeccompare}. The SEDs for M87, Sombrero, and M94 do not reproduce the overall shapes of our spectra well, while the models for M81, NGC~1052, and NGC~4258 follow our observed continuum shapes more closely. These closer continuum matches are associated with models using $\delta_{\rm AGN}=0.0$ and $OA=10^\circ$. In contrast, the M87 and Sombrero models use $\delta_{\rm AGN}=0.25$, while the M94 model has the largest discrepancy, with $\delta_{\rm AGN}=0.5$ and $OA=70^\circ$. All six models use $i=80^\circ$, so inclination does not distinguish their relative agreement with our spectra. Because the remaining parameters also vary together, these comparisons do not isolate the effect of any single parameter.

The silicate features provide a further distinction in \ref{fig:sedspeccompare}. All four SEDs with $\tau_{9.7}=11$ (Sombrero, M94, M81, and NGC~4258) show silicate absorption, whereas the corresponding decomposed AGN spectra show silicate emission. The NGC~1052 SED, with $\tau_{9.7}=3$, shows no silicate feature, while its decomposed AGN spectrum shows emission. The lower optical depth therefore gives a closer representation of the decomposed spectral feature than the models with absorption, but still does not reproduce our observation.

Updated modeling of LLAGN SEDs is clearly necessary, likely with less strict reliance on RIAF models. The inconsistencies we observe between ADAF-based LLAGN SEDs and our decomposed spectra in the IR support previous work that RIAF are insufficient to fully describe LLAGN SEDs and require a significant jet contribution for SED fitting \citep{Prieto2016}. Any future models should incorporate strong jet emission and diffuse dust geometries that produce silicate emission, see \S\ref{subsec:silicate}.

\subsection{LLAGN in IR Color-Color Diagrams} \label{subsec:wisecolor}

The hot dust surrounding traditional AGN produces a strong MIR continuum that is distinguishably redder than that of a star-forming galaxy. This is true for both obscured and unobscured AGN where the emission from the AGN dominates over stellar emission from the host galaxy. Because of this, infrared colors are a common and efficient method for IR AGN selection \citep[e.g.][]{jarrett2011,mateos2012,stern2012, secrest2015}. For lower-luminosity AGN, the dominance of galaxy light can make it hard to detect the AGN, making this method less effective \citep[e.g.][]{satyapal2014,assef2018}, and also making it hard to know whether low-luminosity AGN have infrared colors similar to those of higher-luminosity AGN. Because host galaxy contamination has been removed from our decomposed AGN, we can probe their intrinsic infrared colors for the first time and compare these to typical bright AGN.

\begin{figure}
    \centering
    \includegraphics[width=\linewidth]{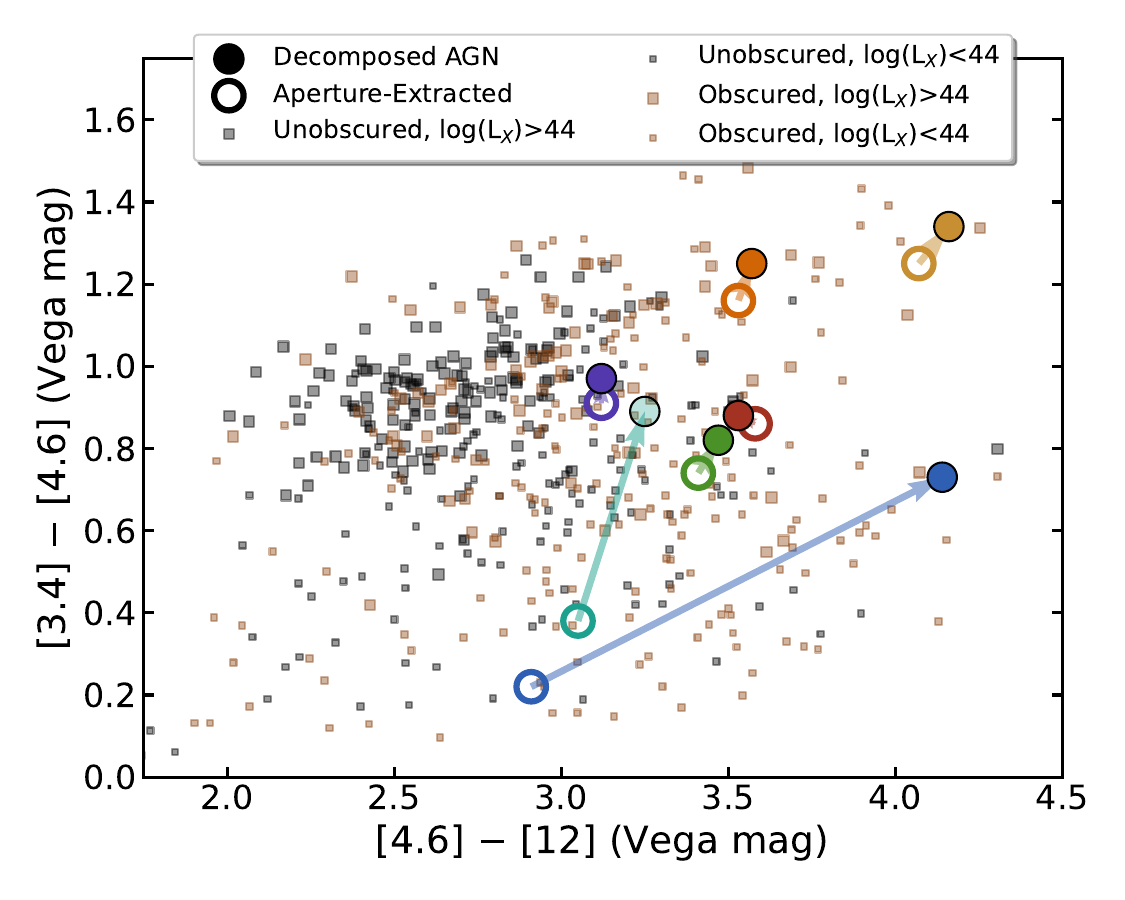}
    \caption{Flux ratio diagram comparing synthetic photometry of our LLAGN (circles) to \textit{WISE} data from 604 nearby AGN from  \citet{ichikawa2017} (squares). Unfilled circles connected to corresponding filled circles show the difference in colors from aperture-extracted spectra to those from our decomposed AGN spectra. For the comparison sample, we define "obscured" AGN shown with brown markers as those with log(N$_\text{H}$) $>22$~[cm$^{-2}$]. Large square markers show brighter AGN with log(L$_\text{14-195~keV})>44$~[erg~s$^{-1}$]. Marker color for each of our seven targets matches those used in Fig.~\ref{fig:spectra}.}
    \label{fig:wisecolor}
\end{figure}

To compare flux ratios in a similar way to the color selection in the commonly used \textit{WISE} bands, we use the python-based synthetic photometry package \textsc{synphot} to calculate Vega magnitudes for \textit{WISE} W1 (3.4~$\mu$m), W2 (4.6~$\mu$m), and W3 (12~$\mu$m) bands from our target spectra. These synthetic magnitudes are calculated for both 75\% encircled energy aperture-extracted spectra and our decomposed AGN spectra to highlight the difference in color with and without host galaxy contribution, even in these nearby galaxies with high spatial resolution observations. We plot synthetic colors for our seven targets in Fig.~\ref{fig:wisecolor}, along with \textit{WISE} colors for 604 nearby AGN from the Swift/BAT Catalog \citep{ichikawa2017}. We note that \textit{WISE} has much lower angular resolution, and measurement by \textit{WISE} of even quite nearby LLAGN would suffer from significant galaxy contamination.

We first examine the impact of galaxy contamination and resolution on this IR flux ratio diagram for our sources. In all cases, the open circles, which are measured from the 75\% encircled energy radii aperture spectra, are bluer in both colors than the unresolved emission, as expected. In particular, for the two faintest and most galaxy-dominated spectra, the \textit{WISE} colors for Sombrero and M94 both lie well below the majority of AGN before isolation. This certainly suggests that as galaxy contributions increase, it becomes increasingly difficult to identify LLAGN. A similar trend is seen in the \textit{WISE} comparison data of nearby galaxies from \citet{ichikawa2017} in Fig.~\ref{fig:wisecolor}, where lower-luminosity sources that are more likely galaxy contaminated show bluer color in both NIR and MIR wavelengths, while luminous AGN are redder. 

To gauge whether the broadband SED has similar colors in LLAGN and brighter AGN, we compare our decomposed LLAGN colors to those from the \citet{ichikawa2017} sample with $\rm log(L_{X})>44$, as their AGN dominated emission similarly contain minimal galaxy contribution. The average [3.4]-[4.6] and [4.6]-[12] colors of this bright subsample are 0.96$\pm$0.014 and 2.9$\pm$0.02 respectively. The average [3.4]-[4.6] color for our unresolved model AGN sample returns a very similar result of 0.99$\pm$0.08. However, the average [4.6]-[12] of our isolated LLAGN sample is 3.60$\pm$0.15. Thus, the decomposed LLAGN nuclei have clearly redder [4.6]-[12] colors than the majority of higher-luminosity AGN. Investigating the cause of this redder MIR color, we find no correlation with silicate emission strength (see~\S\ref{subsec:silicate}), but do see a correlation with the steep slope measured in their continua redward of $\sim8\mu$m, discussed in \S\ref{subsec:slopes}. Sombrero, NGC~4258, and NGC~1052 lie in regions mostly full of obscured sources (with $\rm log(N_{ H}/{\rm cm}^{-2}) > 22$); both NGC~4258 and NGC~1052 fall above this limit, but our other sources have lower obscuration and still show redder MIR colors than the median of the bright AGN sample, suggesting true SED differences.

\section{Conclusions}\label{sec:conclusion}

We used \textit{JWST} NIRSpec and MIRI/MRS spectroscopy of M87, Sombrero, M94, M81, NGC~1052, NGC~4258, and NGC~4395  from the ReveaLLAGN survey to extract clean IR continua. For each target, we decomposed the unresolved AGN spectra to remove contamination from their surrounding host galaxy using 2D brightness profiles.  We present continuous IR spectra for our seven LLAGN, ranging from 1.7~$\mu$m to 24~$\mu$m. Comparison with aperture-extracted spectra demonstrates the effectiveness of our AGN decomposition method in isolating the intrinsic nuclear emission. Stellar CO absorption features at 2.3~$\mu$m and flattened near-infrared continua associated with stellar contamination are fully removed or substantially minimized in our decomposed spectra. Even in the most host-dominated systems, M94 and Sombrero, their resulting spectra recover a clean power-law NIR continuum and residual CO is reduced to only $\rm \sim3\%$ of its original flux. These results confirm that our decomposition technique effectively removes host-galaxy light, enabling accurate broad-spectrum SED fitting and robust characterization of the intrinsic infrared properties of LLAGN. Our decomposed spectra reveal variations and shared characteristics between IR SEDs of LLAGN:

\begin{itemize}

    \item We see broad silicate emission features in our decomposed spectra and present measurements in Table~\ref{tab:silicate}. This strong silicate emission provides strong evidence that warm circumnuclear dust survives on parsec scales, even at very low accretion rates. AGN-to-aperture flux ratios reveal that 9.7~$\mu$m emission is predominantly unresolved, corresponding to maximum spatial scales of $\rm \sim 7-38 pc$. We measure shifts in peak wavelengths of silicate emission features, with the 9.7~$\mu$m feature peaking at longer wavelengths ($\rm>10.4\mu m$) and the 18~$\mu$m feature peaking at shorter wavelengths ($\rm<17.3\mu m$) than Galactic ISM silicates. This is consistent with trends observed in higher-luminosity AGN.
    
    The widespread presence of unresolved silicate emission argues against a significant decrease in circumnuclear dust in LLAGN. It instead favors modified dust geometries like a modified clumpy torus, optically thin remnants of a dissipating torus, or compact dust filaments near the nucleus.
    The lack of silicate absorption in our decomposed spectra disagrees with model LLAGN SEDs from \citet{lopez2024}, which present silicate features at unshifted wavelengths and predominantly in absorption.

    \item We characterize the continuum shapes of our decomposed AGN spectra by fitting power-law functions to NIR (2.8–5.2~$\mu m$) and MIR (13–26~$\mu m$) wavelength ranges. The NIR slopes are broadly consistent with those of higher-luminosity AGN from \cite{brown2019}, supporting previous findings that LLAGN can exhibit IR SED shapes similar to Seyferts. We also find that sources with the steepest NIR continua tend to have the highest $\rm N_H$. In contrast, we measure systematically steeper MIR slopes that distinguish our LLAGN from higher-luminosity AGN. These steep slopes also disagree with LLAGN model spectra from \citet{lopez2024}. Future broad-spectrum SED fits using our decomposed spectra will provide insight into the physical origin of these steep MIR spectra.
 
    \item We test established IR--X-ray relations by comparing integrated 2.2~$\mu m$ and 12~$\mu m$ luminosities of our decomposed LLAGN to their 2–10 keV X-ray luminosities. We find that low-Eddington ratio sources exhibit an enhanced NIR-to-X-ray luminosity ratio relative to higher-Eddington AGN, supporting previous findings from \citet{dumont2020}. We measure a correlation between NIR-to-X-ray luminosity ratio and radio radio-loudness, consistent with a jet origin for the observed excess. However, we do not see a significant 12~$\mu m$ excess in our low-Eddington sample. Our targets also align well with established linear $\rm L_{12~\mu m}$--$\rm L_{X}$ relations despite the physical mechanism driving the correlation likely differing from that of higher-luminosity AGN.

    \item We use synthetic photometry to calculate IR colors from both aperture-extracted and decomposed AGN spectra. We confirm that host galaxy contamination biases measurements towards bluer colors, highlighting the importance of decomposition for future observations of LLAGN. Similar to our results with the power-law slopes, the synthetic photometry from our decomposed spectra exhibits [3.4]–[4.6] colors comparable to those of luminous AGN from \cite{ichikawa2017}, but shows systematically redder [4.6]–[12] colors. These redder MIR colors are not correlated with silicate emission strength, but instead track the steeper MIR continuum redward of $\rm \sim8\mu m$, indicating that their unique IR colors represent a genuine difference in SEDs of LLAGN.
    
\end{itemize}

Upcoming papers (Prieto et al., {\em in prep} and Duncan et al., {\em in prep}) will fit physical models to the decomposed LLAGN spectra presented here. With $\sim20\%$ variability expected in the AGN-dominated emission of our decomposed spectra, repeat \textit{JWST} observations of these targets would be highly informative.

\section*{Acknowledgements}
We would like to thank I. E. L{\'o}pez and O. Gonz{\'a}lez-Mart{\'\i}n for their support, insightful discussion, and willingness to share data. We also thank the anonymous referee for their useful comments that helped improve the paper. D.O., A.S., and K.G. were all supported by James Webb Space Telescope general observer grant JWST-GO-02016. S.M. and R.D. are supported by a European Research Council (ERC) Synergy Grant "BlackHolistic" grant No. 10107164. R.M.P. acknowledges support from NASA under award No 80NSSC23M0104. L.C.H. was supported by the National Science Foundation of China (12233001) and the China Manned Space Program (CMS-CSST-2025-A09). 
N.A. and J.A.F.O. acknowledge financial support by the Spanish Ministry of Science, Innovation and Universities (MICIU/AEI
/10.13039/501100011033), by ``ERDF A way of making Europe'' and by ``European Union NextGenerationEU/PRTR'' through the grants PID2021-124918NB-C44 and CNS2023-145339; MCIN and the European Union -- NextGenerationEU through the Recovery and Resilience Facility project ICTS-MRR-2021-03-CEFCA; MICIU and ERDF/EU through the grant PID2023-147386NB-I00. A.F.K. acknowledges funding from the Austrian Science Fund (FWF) [grant DOI 10.55776/ESP542].
\section*{Data Availability}

The data underlying this article are available in the article and in its online supplementary material.


\bibliographystyle{mnras}
\bibliography{bibliography}



\appendix 
            
\section{Target Properties and Data Processing} \label{appendix:data}

\subsection{\textit{JWST} Processing Pipeline Details}\label{appendix:pipeline}

Two data processing choices are of particular importance for our methods: we align all processed data to the IFU coordinate system as a common frame of reference and apply pixel-to-pixel background subtraction. Using sky-alignment would result in different rotation angles between targets, so IFU-alignment allows to fit models in \S\ref{sec:fit} without additional rotation or convolution.

\begin{figure}
    \centering
    \includegraphics[width=\linewidth]{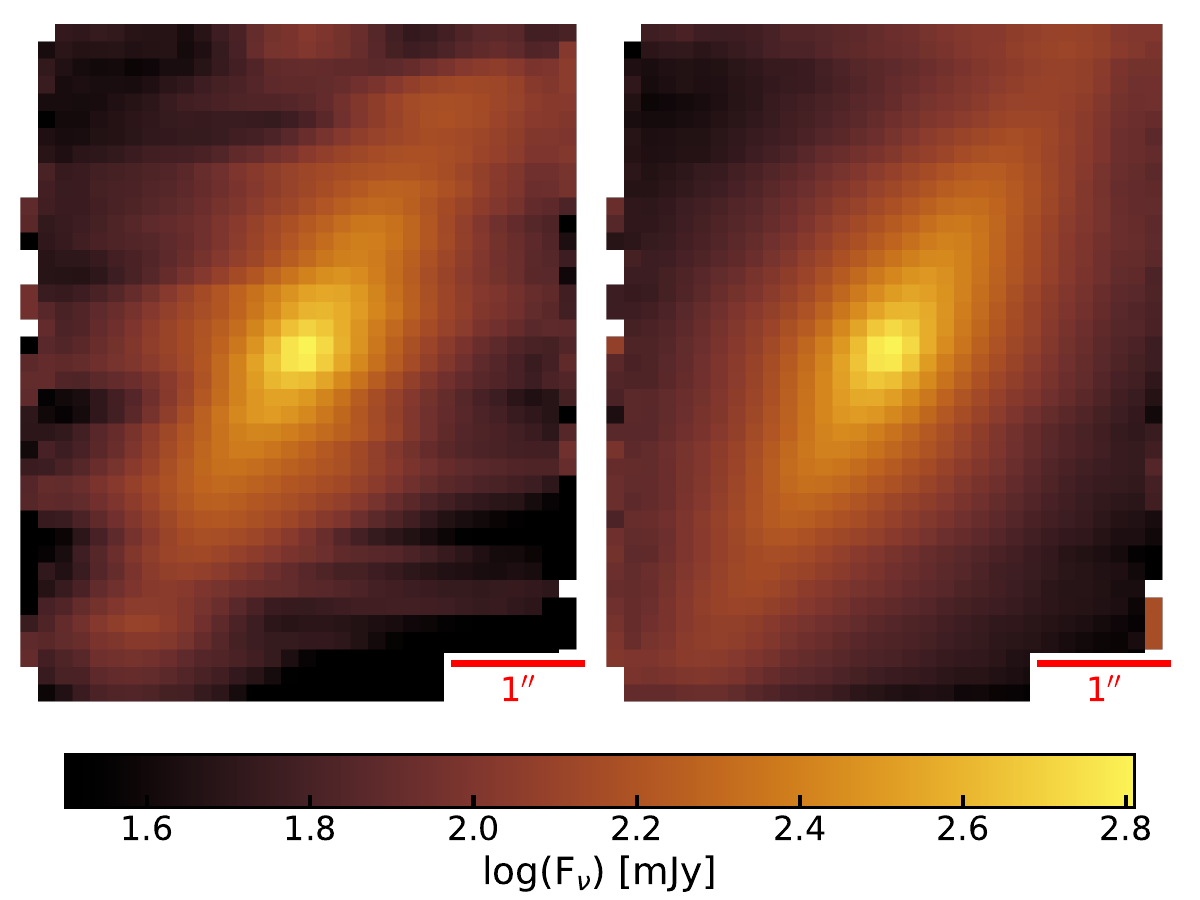}
    \caption{Comparison of resulting data from different \textit{JWST} pipeline background subtraction methods. Both images show the spectral median 2D image of Ch1 MIRI data for Sombrero. {\emph Left} - data cube processed using the default "Master" background subtraction. {\emph Right} - data cube processed using pixel-to-pixel background subtraction. Both images show log fluxes with the same color range.}
    \label{fig:stripes}
\end{figure}

The \textit{JWST} pipeline utilizes master background subtraction by default. This background subtraction method did not remove the significant horizontal striping correlated with 1/f noise, and resultant imaging led to less accurate 2D fits. The use of pixel-to-pixel background subtraction was a significant improvement, as it eliminated 1/f noise striping and resulted in overall clearer morphology, as shown in Fig.~\ref{fig:stripes}.
In this process, we found that combining MIRI/MRS backgrounds from multiple observations increased the S/N for some of our science observations; in particular for M81, M94, NGC~4258, and NGC~4395 we used a combined background, while individual backgrounds performed better in the other three galaxies.
The resulting cleaner 2D imaging yields higher accuracy and reduced residuals from our brightness-profile modeling methods, as described in \S\ref{sec:fit}.

\subsection{Eddington Ratio Calculation}\label{appendix:eddington}

For consistency, we calculate Eddington ratio for ReveaLLAGN targets and comparison samples following the methods of \citet{goold2026}. We first define Eddington luminosity as:

\begin{equation}\label{eq:ledd}
L_{\mathrm{Edd}} = 1.26 \times 10^{38} \left( \frac{M_{\mathrm{BH}}}{M_\odot} \right)~\mathrm{erg\,s^{-1}}, 
\end{equation}.

We calculate a 2-10~keV X-ray-based bolometric correction using methods from \citet{duras2020}, defined as:
\begin{equation}
K_{\rm{X}}(L_{\rm{2-10keV}}) = a \left[ 1 + \left( \frac{\log \left( \frac{L_{\rm{2-10keV}}}{L_\odot} \right)}{b} \right)^c \right],
\label{eq:Kx_Lx}
\end{equation}

where \(a = 15.33\), \(b = 11.48\), and \(c = 16.2\).\\

Using eq.~\ref{eq:Kx_Lx}, we define bolometric luminosity as:
\begin{equation}\label{eq:lbol}
  L_{\rm{Bol}} = K_{\rm{X}}(L_{\rm{2-10~keV}}) \times L_{\rm{2-10~keV}}.
\end{equation}

Because Eq.~\ref{eq:Kx_Lx} flattens at low luminosities, the seven ReveaLLAGN targets have a roughly constant K$_\text{X}\sim$15.34. This value lies within the range of corrections calculated or applied for LLAGN \citep[e.g.][]{ho2009,lopez2024,gupta2025,gupta2024} and allows methodological consistency with calculated Eddington ratios for higher-luminosity comparison data.

\section{Fitting Quality}\label{appendix:mads}

\begin{figure}
    \centering
    \includegraphics[width=\linewidth]{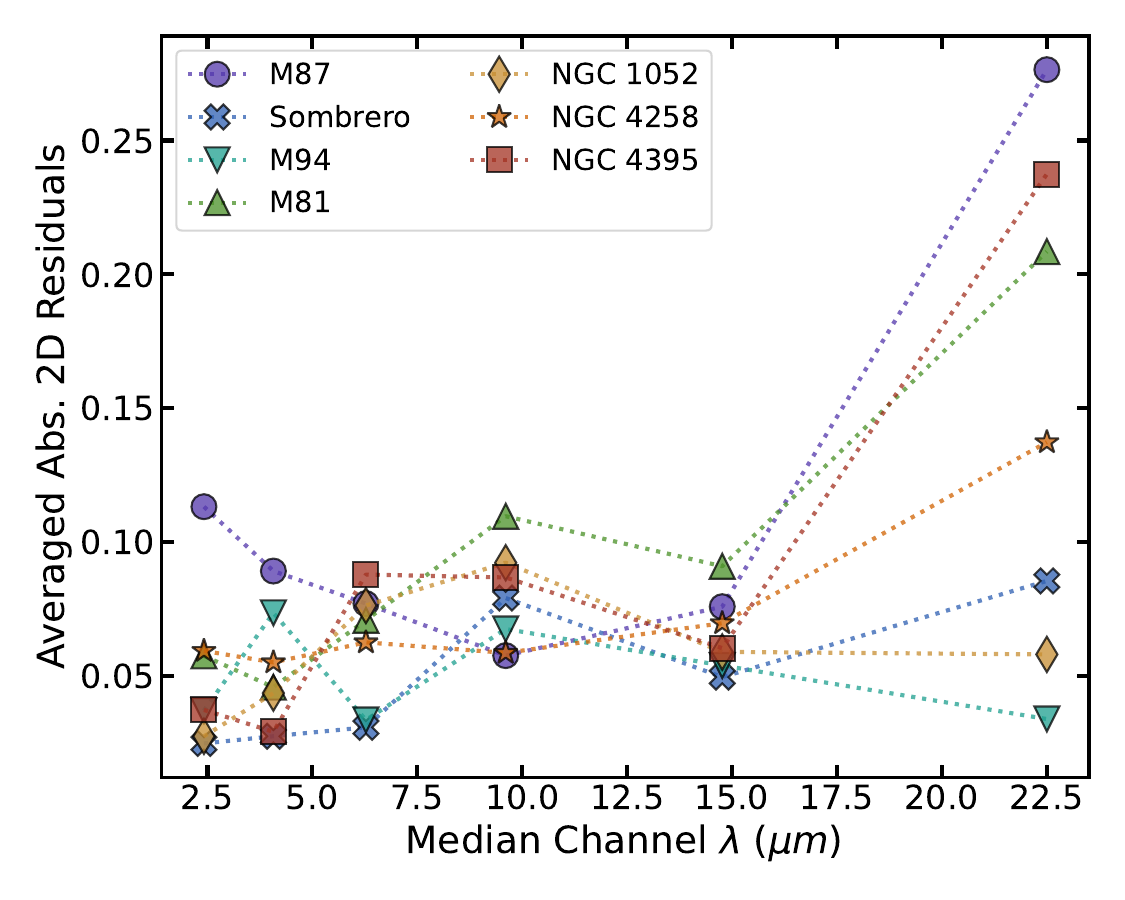}
    \caption{Average fractional residuals for our 2D fits across the full wavelength range. Each point represents the median fractional pixel-to-pixel residual in individual wavelength slices of a given channel and is plotted at its corresponding central wavelength. The figure shows that the majority of our fits overall lie within $10-15\%$ of the observed flux, with discrepancy increasing with wavelength. The figure also depicts our reason to consider our modeled AGN spectra $\lesssim24~\mu m$, since MIRI Ch4 data causes increasingly poor fits above this wavelength.}
    \label{fig:mads}
\end{figure}

We can assess the quality of our fits using the residuals in the individual wavelength slice fits. We first calculate the median absolute deviation (MAD) of the fractional residuals within a 1.7'' aperture. We combine these residuals into a single assessment in each channel by averaging the MAD in each channel for all galaxies as shown in Fig.~\ref{fig:mads}. This shows that the typical residuals increase with wavelength, and are below 10\% pixel-to-pixel for most galaxies shortwards of 10$\mu$m. They are largest in MIRI channel 4 due to its high and variable background, with residuals approaching 30\%.

\section*{Authors' affiliations}
\noindent
$^{1}$Department of Physics \& Astronomy, University of Utah, E2108 Stewart Building 270 S 1400 E, Salt Lake City, UT 84112, USA\\
$^{2}$Centro de Estudios de F\'{i}sica del Cosmos de Arag\'{o}n (CEFCA), Plaza San Juan 1, E–44001, Teruel, Spain\\
$^{3}$European Space Agency, c/o STScI, 3700 San Martin Drive, Baltimore, MD 21218, USA\\
$^{4}$Max-Planck-Institut f{\"u}r Astronomie, K{\"o}nigstuhl 17, D-69117, Heidelberg, Germany\\
$^{5}$Anton Pannekoek Institute for Astronomy, University of Amsterdam, Science Park 904, 1098XH Amsterdam, The Netherlands\\
$^{6}$Gravitation Astroparticle Physics Amsterdam (GRAPPA) Institute, University of Amsterdam, Science Park 904, 1098XH Amsterdam, The Netherlands\\
$^{7}$Department of Astronomy and Astrophysics and Institute for Gravitation and the Cosmos, Penn State University, 525 Davey Lab, 251 Pollock Road, University Park, PA 16802, USA\\
$^{8}$Department of Astrophysics, University of Vienna, T\"urkenschanzstra{\ss}e 17, 1180 Vienna, Austria\\
$^{9}$Department of Astronomy, University of Michigan, 1085 S. University Ave., Ann Arbor, MI 48109, USA\\
$^{10}$Department of Astrophysical Sciences, Princeton University, Princeton, NJ 08544, USA\\
$^{11}$Kavli Institute for Astronomy and Astrophysics, Peking University, Beijing 100871, China\\
$^{12}$Department of Astronomy, School of Physics, Peking University, Beijing 100871, China\\
$^{13}$Department of Physics \& Astronomy, Vanderbilt University, Nashville, TN 37235, USA\\
$^{14}$Department of Physics, University of Nevada, Reno, NV 89557, USA\\
$^{15}$Nevada Center for Astrophysics, University of Nevada, Las Vegas, NV 89154, USA\\
$^{16}$Universidad de La Laguna (ULL), Dpto. Astrof\'{i}sica, Avd. Astrof\'{i}sico Fco. S\'{a}nchez s/n, 38206 La Laguna, Tenerife, Spain\\
$^{17}$Instituto de Astrof\'{i}sica de Canarias (IAC), C/V\'{i}a L\'{a}ctea s/n, 38205 La Laguna, Tenerife, Spain\\
$^{18}$Universit{\"a}ts-Sternwarte, Fakult{\"a}t f{\"u}r Physik, Ludwig-Maximilians-Universit{\"a}t M{\"u}nchen, 81679 M{\"u}nchen, Germany\\
$^{19}$George Mason University, Department of Physics and Astronomy, MS3F3, 4400 University Drive, Fairfax, VA 22030, USA\\
$^{20}$George P. and Cynthia W. Mitchell Institute for Fundamental Physics and Astronomy, Department of Physics \& Astronomy, Texas A\&M University, 4242 TAMU, College Station, TX 77843, USA\\
$^{21}$Institute of Astronomy, University of Cambridge, Madingley Road, CB3 0HA, UK

\bsp	
\label{lastpage}
\end{document}